\documentclass[11pt]{article}
\usepackage{amssymb}
\usepackage{amssymb}
\usepackage{amsmath}
\usepackage{epsfig}
\title{Notes on the Verlinde formula in non-rational conformal field
  theories}
 \author{Charles Jego${}^1$ and Jan Troost${}^2$ \\ ${}^1$
  Centre de Physique Th\'eorique de l'Ecole
  Polytechnique\thanks{Unit{\'e} mixte
    de recherche du CNRS, UMR 7644. Preprint: CPHT RR014.0306}\\ 91128
  Palaiseau Cedex, France \\  ${}^2$ Laboratoire de Physique
  Th\'eorique de l'\'Ecole Normale Sup\'erieure\thanks{Unit{\'e} mixte
  du CNRS et de l'Ecole Normale Sup{\'e}rieure, UMR 8549. Preprint:
  LPTENS-06/01}  \\ 24, Rue Lhomond  75231 Paris Cedex  05, France
  $\ $ \\
  E-mail:  Charles.Jego(AT)cpht.polytechnique.fr, Jan.Troost(AT)lpt.ens.fr  \\
}
\date{}

\begin{document}
\maketitle

\abstract{We investigate to which extent the Verlinde formula can be
  expected to remain valid in non-rational conformal field
  theories. Moreover, we check the validity of a proposed generic
  formula for localized brane one-point functions in non-rational
  conformal field theories.}

\section{Introduction}

It is fair to say that we have acquired a systematic understanding of
unitary rational conformal field theories (see
e.g.\cite{DiFrancesco:1997nk}). We have solved many classification
problems (of which the classification of minimal models and $SU(2)$
modular invariant partition functions are simple examples
\cite{Cappelli:1987xt}), and have laid bare a lot of the algebraic
structure that underlies them (e.g.their integrable highest weight
representations, characters, simple currents, etc.). For non-unitary
rational conformal field theories (i.e.conformal field theories which
have a finite number of primaries with respect to their chiral
algebra, but which are not necessarily unitary), our understanding has
advanced less, but partial results are known. In particular, in all
these theories a lot is known about the relation between the modular
data, and the fusion of representations as encoded in the operator
product expansions (see e.g.\cite{Gannon:2001uy}). An important
relation is given by the Verlinde formula, which encodes the fact that
the modular S-matrix diagonalizes the fusion matrix
\cite{Verlinde:1988sn}.

In these notes, we take a step in the analysis of non-rational
conformal field theories along similar lines, and we investigate which
algebraic structures that we have discovered in rational conformal
field theories can be extended to the non-rational case. The solution
of non-rational conformal field theories, like their rational cousins,
can often usefully be attacked by identifying special algebraic
properties (null-vectors) of the representation spaces, that are next
exploited in a differential calculus that may lead to a solution for,
say, bulk or boundary three-point functions. While in the rational
case one has developed in parallel the algebraic approach to these
conformal field theories (identifying characters, their modular
transformation properties, their modular invariant combinations, and
their relation to the fusion algebra), for the non-rational case, this
problem has been attacked less. Although many results in this spirit
are known (see e.g.
\cite{Zamolodchikov:2001ah,Fateev:2000ik,Teschner:2000md,Eguchi:2003ik,Israel:2004jt,Fotopoulos:2004ut,Ahn:2004qb}),
we believe it is useful to review and supplement them in these notes
and in particular in the light of the possibility of extending the
Verlinde formula to a subsector of non-rational conformal field
theories.  In rational conformal field theories, the Verlinde formula
leads to the construction of a set of boundary states having a
reasonable boundary spectrum (namely Cardy states \cite{Cardy:1989ir},
defined through the Cardy formula). These states have found many
applications in string theory as describing non-perturbative states
carrying open string excitations. In non-rational conformal field
theories as well, the analogue of the Verlinde formula allows for an
efficient construction of a subset of boundary states, directly from
the modular data \cite{Teschner:2000md,Eguchi:2003ik}. Thus, a
systematic analysis of the Verlinde formula should be useful in
constructing D-branes in (non-trivial non-compact) string theory
backgrounds.

Our paper is structured as follows: we first very briefly remind the
reader of properties of the Verlinde formula and modular S-matrices in
rational conformal field theory. Next, we discuss in detail in which
subsectors of bosonic Liouville, $N=2$ Liouville theory (i.e. the
supersymmetric $SL(2,R)/U(1)$ coset) and the $H_3^+$ conformal field
theory we can find a form of the Verlinde formula. We conclude in the
final section with an attempt to delineate our generic expectation for
the domain of validity of the Verlinde formula in non-rational
conformal field theories.

In two appendices, we show in detail how to find the fusion formulas
for degenerate representations and non-degenerate representations
for both bosonic Liouville theory and
the $H_3^+$ model from the corresponding generic three-point functions
for representations in the unitary spectrum by analytic continuation
and a careful analysis of the analytic structure of the operator
product expansions.

\section{The Verlinde formula in rational conformal field theories}
In this section we recall a few salient features of 
the algebraic structure of rational conformal field theory, as a point of
reference for our future treatment of the non-rational case. The section will
be a review of well-known facts.
\subsection*{The unitary rational case}
The context in which we have firstly developed our understanding of
generic algebraic structures underlying conformal field theories is
the case of unitary rational theories. We briefly review part of its
solid structure (see e.g. \cite{Gannon:2001uy,Fuchs:1993et} for
nice expositions). These theories are based on a finite number of
primary fields (i.e. irreducible modules of the (chiral) vertex
operator algebra). The associated irreducible representations of the
chiral algebra have characters:
\begin{eqnarray}
\chi_a (\tau) &=& q^{-c/24} Tr_{ {\cal H}_a} q^{L_0},
\end{eqnarray}
where $\chi_a(\tau)$ is the character associated to the representation
with Hilbert space ${\cal H}_a$ (corresponding to the primary $a$
which takes values in the finite set of primaries $P$). The modular parameter
is $q=e^{2 \pi i \tau}$, the central charge of the CFT is $c$ and
$L_0$ is the scaling operator in the two-dimensional conformal
algebra. (In general, we would allow for extra variables in the
characters, keeping track of more quantum numbers in the case where
the chiral algebra is extended.) The characters yield a representation
of the modular group $SL(2,\mathbb{Z})$ which is generated by the $S$
and $T$ transformations. Under these transformations, these characters
transform amongst themselves as:
\begin{eqnarray}
\chi_a(-1/\tau) = \sum_{b \in P} {S_{a}}^b \chi_b(\tau) \ ,
\hspace{1cm} \chi_a(\tau+1) = \sum_{b \in P} {T_a}^b \chi_b(\tau).
\end{eqnarray}
The representation of the modular group generated by these matrices,
the modular data, satisfy many powerful relations in the case of
(unitary) rational conformal field theories. These modular data
dominate the analysis of the CFT on the torus (i.e. the genus zero
vacuum integrand in string theory) and on the disc with one puncture
as well as on the annulus (i.e. the closed string one-point function
and the one-loop open string vacuum integrand).

In unitary rational conformal field theories, there is an identity
field (with label $0$), the modular S-matrix is unitary and symmetric, the
matrix $T$ is diagonal and of finite order, and they satisfy more
non-trivial properties. The one we will concentrate on in these notes
is the relation between the modular S-matrix and the structure
constants of the fusion ring, i.e. the most commonly encountered case
of the Verlinde formula:
\begin{eqnarray}
{{\mathcal{N}}_{ab}}^c = \sum_{d \in P} \frac{ {S_a}^d {S_b}^d
  {({S^{-1}})_d}^c}{ {S_0}^d} \label{verlindercft}
\end{eqnarray}
The numbers ${{\mathcal{N}}_{ab}}^c$ are positive integers, and they
encode information on the multiplicity of the operator product
expansion of two chiral primaries (or on the three-point function on
the sphere). The fusion matrices $N_a$, defined by ${(N_a)_b}^c =
{{\mathcal{N}}_{ab}}^c$, are diagonalized by the modular S-matrix, and
the eigenvalues are ${S_a}^d / {S_0}^d$.

Related to the definition of these modular data is the study of
modular invariants (which correspond to torus amplitudes in string
theory), as well as the study of non-negative integer representations
of the fusion ring (NIM-reps), i.e. of (non-negative integer) matrices
$X_a$ that satisfy $X_a X_b = {\mathcal{N}_{ab}}^c X_c$. That
systematic study has advanced considerably since the advent of
rational conformal field theory.

Many representations of these modular data and of the fusion ring are
known.\footnote{For instance, finite groups represent modular data
  \cite{Gannon:2001uy}.} We will only briefly remind the reader of how
these relations work in a few examples of unitary rational conformal
field theories. The examples will serve as points of comparison for
our later treatment of the non-rational case.

\subsection*{A few examples}

As a reminder, we briefly show how the algebraic structure
is realized  in a few examples.
For a $U(1)$ boson at integer radius-squared $R=\sqrt{k \alpha'}$, we
have an extended chiral algebra including momentum operators, and the
primaries are labeled by $n \in \mathbb{Z}_{2k}$. We have that the
modular S-matrix and fusion rules are given by:
\begin{eqnarray}
{S_n}^{n'} &=& \frac{1}{\sqrt{2k}} e^{- i \pi \frac{n n'}{k}} \\
{{\mathcal{N}}_{nn'}}^{n''} &=& \delta_{n+n',n''}^{2k}
\end{eqnarray}
The upper index of the delta-symbol indicates its periodicity. The
Verlinde formula (\ref{verlindercft}) is easily checked to hold. This example can be
extended to conformal field theories on tori, associated to generic
lattices (i.e. string theory on tori) \cite{Gannon:2001uy}.

The next example is the $SU(2)_{k-2}$ Wess-Zumino-Witten model.
Chiral primaries are labeled by $j=0,\frac{1}{2}, 1, \dots , \frac{k-2}{2}$.
The modular S-matrix is:
\begin{eqnarray}
{S_j}^{j'} &=& \sqrt{\frac{2}{k}} \sin \left( \frac{\pi
  (2j+1)(2j'+1)}{k} \right)
\end{eqnarray}
while the fusion rules are:
\begin{eqnarray}
{\mathcal{N}_{j j'}}^{j''} = \left\{ \begin{array}{ll}
1 \ \ \text{if} \ \left\{
     \begin{array}{cc}
|j-j'| \le j'' \le \text{min} (j+j',k-2-j-j') \\
j+j'+j'' \in \mathbb{Z}
\end{array} \right. \\
0 \ \ \text{else}
\end{array} \right.
\end{eqnarray}

Again, the Verlinde formula can be checked straightforwardly
\cite{Verlinde:1988sn} using $SU(2)$ group character formulas. The
above data are again a realization of unitary rational modular data.
The structure extends to generic WZW models, i.e. generic affine
Kac-Moody algebras.  More generally, the validity of the Verlinde
formula has been found for the minimal models, for unitary coset
models (parafermions), for their $N=1$ and $N=2$ supersymmetric
cousins, and it has manifestations in many other contexts.  Some
systematic understanding of the Verlinde formula was acquired on the
basis of the axioms of conformal field theory \cite{Moore:1988qv}, but
a mathematical proof of the Verlinde formula for a large class of
rational conformal field theories based on a minimal set of
assumptions has only recently been provided \cite{Huang:2004dg}.


\subsection*{The non-unitary rational case}

It is interesting to briefly take note of what happens in the
non-unitary  rational conformal field theories\footnote{These non-unitary
  theories include examples immediately relevant to known physics of
  two-dimensional critical systems, e.g. the Yang-Lee singularity.}. 
An interesting set of
generic comments was made in \cite{Gannon:2003de}. Most striking is
the fact that one can argue for the identification of a special
primary that not necessarily corresponds to the vacuum (but that can
be seen as a unity). The properties of the usual vacuum/identity field
in the unitary case are now shared by the vacuum and the new unity
field. One can characterize it, for instance, by demanding that it
corresponds to a positive column in the S-matrix. In particular, it
takes over the role of the identity in the calculation of the Verlinde
formula.  A good example in this class is the $SU(2)$ conformal field
theory, at (possibly negative) fractional level. Already in this
example, fusion, the  Verlinde formula, positivity of the fusion ring
structure constants, the identification of the relevant vertex
operator algebra to define the ring, and its realization in conformal
field theory are far less trivial, and have not been entirely settled
(see e.g.
\cite{Awata:1992sm,Ramgoolam:1993gt,Feigin:1993dt,Gaberdiel:2001ny,Lesage:2002ch}).

Let's now turn to the central subject of these notes, the analysis of the
Verlinde formula for non-rational conformal field theories.

\section{Non-rational conformal field theories and the Verlinde formula}

We briefly reviewed what we know to be a generic structure (modular
data) dictating the allowed modular invariants, NIM-reps, and the
closed string one-loop amplitudes, disc one-point functions and their
associated D-branes in rational conformal field theories
\cite{Cardy:1989ir}. For non-rational conformal field theories, our
approach at this stage is less systematic. We first review the list of
examples of non-rational conformal field theories where we have a
reasonable handle on the relevant characters, the modular S-matrices,
and the brane spectrum. From it, we will extract generic lessons,
point out the differences with the rational cases (both unitary and
non-unitary), and delineate what generic systematics to expect.

One should compare this to the modular bootstrap approach for building
branes -- implicitly this makes use of the existence of an analogue of
the Verlinde formula. We believe it is useful to shed a different
light on some of these calculations, since an algebraic understanding
of non-rational CFT is a prerequisite for a better understanding of
the modular bootstrap. In the following, we will work through the
example of bosonic Liouville theory, the $H_3^+$ model and the
$SL(2,\mathbb{R})/U(1)$ coset.


\subsection{Bosonic Liouville}
To set up an analogue of modular data and a Verlinde-type formula in
bosonic Liouville theory \cite{Zamolodchikov:2001ah}, we review the
characters of the Virasoro algebra, their modular transformation
properties, and some Cardy-type calculations for branes in Liouville
theory, in a form suitable for interpretation in terms of a Verlinde
formula. We will supplement the review with remarks which turn out to
generalize to other cases to be discussed later on.

The central charge will be defined by:
\begin{eqnarray}
c = 1+ 6 Q^2 \ , \hspace{1cm} Q = b + b^{-1}
\end{eqnarray}
where $b$ is strictly positive and $b^2$ is non-rational. 
Other formulas in Liouville theory are collected in
Appendix~\ref{appliouville}. For non-degenerate representations, we
define a momentum $2 \alpha = Q + \imath s$ with $s \in \mathbb{R}_+$,
and the character and conformal dimension of these representations
are\footnote{Our notations throughout these notes will be $\tau \in H$
  where $H = \{ z \in \mathbb{C} | \Im z >0 \}$ is the upper
  half-plane, $q=e^{2 \imath \pi \tau}$, $\tau' = -1/\tau$ and $q' =
  e^{2 \imath \pi \tau'}$.}:
\begin{eqnarray}
\label{chis}
\label{deltas}
\chi_s (\tau) = \frac{q^{s^2/4}}{\eta(\tau)} \ , \hspace{1cm}
h_s = \frac{1}{4} \left( Q^2 + s^2 \right)
\end{eqnarray}
where $\eta$ is the Dedekind function. For degenerate representations
(which for non rational $b$ have a single null vector at level $n
m$), we take $2 \alpha = Q -  \frac{m}{b} - n b$ with $m$ and $n$
strictly positive integers, and the character and conformal dimension
are:
\begin{eqnarray}
\chi_{m,n} (\tau) = \frac{q^{-(m/b+nb)^2/4} - q^{-(m/b-nb)^2/4}  
}{\eta(\tau)} \ , \hspace{1cm} h_{m,n} = \frac{1}{4} \left( Q^2 - (m/b+nb)^2 \right)
\end{eqnarray}
The modular transformations of the characters are:
\begin{eqnarray}
\label{smss}
\chi_s \left( - \frac{1}{\tau} \right) &=& \int_{0}^{\infty}
{S_s}^{s'} \chi_{s'}(\tau) d s' \ , \hspace{.8cm} {S_s}^{s'} = \sqrt{2}
\cos \left( \pi s s' \right) \\
\chi_{m,n} \left( - \frac{1}{\tau} \right) &=& \int_0^{\infty}
{S_{m,n}}^{s'} \chi_{s'} (\tau) ds' \ , \hspace{.1cm} {S_{m,n}}^{s'} =
2 \sqrt{2} \sinh \left( \pi m \frac{s'}{b} \right) \sinh \left( \pi n
  b s' \right)
\label{degmod}
\end{eqnarray}
Note that the calculation of the modular transformation of the continuous
characters is quite robust, in the sense that many contours, for instance
those parallel to the real axis (instead of the one on the real axis)
 would yield the same result -- there are no poles to be picked up, and the
convergence of the characters ($Im(\tau)>0$) makes the calculation
robust.

A useful addition to the usual discussion of these modular
transformation properties is the check that the modular S-matrix
squares to the identity.  In the non-degenerate sector, this follows
from standard formulas in the theory of cosine Fourier transforms. In
the degenerate sector, the proof is slightly  more subtle, and it is a
useful foreshadowing of the techniques  used later on. Indeed, let us
compute the modular transform of the first equation in (\ref{degmod}):
\begin{eqnarray}
\label{inversion}
\chi_{m,n} (\tau) &=& 2 \int_0^{\infty} ds' 
\sinh \left( \pi m \frac{s'}{b} \right) \sinh \left( \pi n b s'
\right) \int_{-\infty}^{+\infty} d t e^{\imath \pi t s'} \chi_t(\tau)
\end{eqnarray}
We unfolded the integral over real momentum $t$. It is important to
note that we cannot interchange the $t$ and the $s$ integral, since
the $s'$ integral is then divergent. However, we can shift the $t$
contour off the real axis, not encountering any poles, and keeping the
convergence of the integral, to render the $s'$ integral finite after
exchange of the order of integrations (see \cite{Teschner:2000md} for
a similar manipulation). To that end, we need to add a positive
imaginary part to the $t$ integration variable that is larger than
$m/b + n b$, e.g. $\Delta = m/b + n b + \epsilon$ where $\epsilon$ is a
positive number. We can then exchange the integrals and perform the
$s'$ integral:
\begin{eqnarray}
\label{squares}
\chi_{m,n} (\tau) &=&  \int_{-\infty+\imath \Delta}^{+\infty+\imath
  \Delta} d t  \ \chi_t(\tau) \sum_{\epsilon_1,\epsilon_2 \in \{ -1,+1
  \} } - \frac{\epsilon_1 \epsilon_2}{\epsilon_1 2 \pi m/b +
  \epsilon_2 2 \pi n b + 2 \imath  \pi t}
\end{eqnarray}
We then shift the $t$-integral back to the real axis, pick up the poles in the
upper half plane, and find (for $m/b-nb>0$):
\begin{eqnarray}
\chi_{m,n} (\tau) = \chi_{\imath (m/b + n b)} (\tau) - \chi_{\imath
  (m/b-nb)} (\tau)
\end{eqnarray}
This proves that indeed, the modular S-matrix squares to the identity, albeit
in a seemingly roundabout fashion. We will revisit later on this
technique of unfolding the integral over real momentum and shifting
the integration contour in order to invert a modular S-matrix and find
an analogue of the Verlinde formula.

We move on to a discussion of the one-point functions, their precise
relation to the modular S-matrix, the reflection amplitude, and the
fusion rules, and to which extent we can generalize Verlinde's
formula. We will also recall the calculation relating boundary state
partition function to bulk channel exchange. As in rational conformal
field theory, the boundary states relate to the modular S-matrix in
the bulk, while the boundary partition function encodes fusion
coefficients. We revisit the analysis of boundary states in Liouville
theory and make this relation manifest. Firstly, we recall the
calculation of the ZZ-brane spectrum, then we turn to the case of the
FZZT-branes.

\subsection*{Degenerate representations}
The degenerate field $m=n=1$ will play the role of the identity field
in the Verlinde formula. We recall that the wave function for a
ZZ-brane is \cite{Zamolodchikov:2001ah}:
\begin{eqnarray}
\psi_{m,n} (s) &=& \psi_{1,1} (s) \frac{\sinh \left( \pi m
  \frac{s}{b} \right)  \sinh \left( \pi n b s \right)}{\sinh \left(
  \pi \frac{s}{b} \right) \sinh \left( \pi  b s \right)}
\end{eqnarray}
where we used the expression for the one-point function of the $(1,1)$ brane:
\begin{eqnarray}
\psi_{1,1} (s) &=& 2^{3/4} \frac{\imath s}{\Gamma(1 - \imath b s)
  \Gamma(1 - \imath s/b)} (\pi \mu \gamma (b^2))^{- \imath s /(2 b)}
\end{eqnarray}
We note that we have the relation:
\begin{eqnarray}
\psi_{1,1} (s) \psi_{1,1} (-s) = {S_{1,1}}^s
\end{eqnarray}
We want to relate the one-point function for the ZZ-brane to the
modular S-matrix. In order to do that, we recall the expression for
the  Liouville reflection amplitude \cite{Zamolodchikov:1996xm}:
\begin{eqnarray}
\mathcal{R}_L (s) = \frac{\psi_{1,1} (s)}{\psi_{1,1} (-s)} = - \left(
  \pi \mu \gamma (b^2) \right)^{-\imath s/b} \frac{\Gamma(1+\imath b
  s) \Gamma(1+\imath s/b)}{\Gamma(1-\imath b s) \Gamma(1-\imath s/b)}
\end{eqnarray}
which  satisfies $\mathcal{R}_L (s) \mathcal{R}_L (-s) = 1$ and:
\begin{eqnarray}
\psi_{m,n} (s) = \mathcal{R}_L (s) \psi_{m,n} (-s)
\label{refl}
\end{eqnarray}
The wave function and the modular $S$-matrix can then be related as follows:
\begin{eqnarray}
\label{wavefct1}
\psi_{m,n} (s) \sqrt{\mathcal{R}_L (-s)} =
\frac{{S_{m,n}}^s}{\sqrt{{S_{1,1}}^s}}
\end{eqnarray}
Note that in this relation one introduces the reflection coefficient
\cite{Teschner:2000md}, in contrast to the case of the Cardy formula
in rational conformal field theory.  The partition function for two
ZZ-branes calculated in \cite{Zamolodchikov:2001ah} can now be
expressed as:
\begin{eqnarray}
\label{zmn}
Z_{m,n ; m',n'} (\tau) &=& \int_0^{\infty} \psi_{m,n} (s)
\psi_{m',n'} (-s) \chi_s (\tau) ds = \int_0^{\infty} \frac{{S_{m,n}}^s
  {S_{m',n'}}^s}{{S_{1,1}}^s} \chi_s (\tau) ds \\
&=& \sum_{k=0}^{\text{min}(m,m')-1} \ \sum_{l=0}^{\text{min}(n,n')-1}
\chi_{m+m'-2k-1,n+n'-2l-1} (-1/\tau) \nonumber
\end{eqnarray}
where we used the following relation (and its analogue for $b \rightarrow
b^{-1}$):
\begin{eqnarray}
\sinh \left( \pi n b s \right) \sinh \left( \pi n' b s \right) &=&
\sum_{l=0}^{\text{min}(n,n')-1} \sinh \left( \pi b s \right) \sinh
\left( \pi (n+n'-2l-1) b s \right) \nonumber
\end{eqnarray}

Now, in appendix \ref{appliouville} we analyze how to recuperate the
fusion of degenerate representations from the fusion of the
non-degenerate ones that make up the spectrum of the unitary Liouville
conformal field theory, through analytic continuation. We provide many details of the calculation in
Appendix \ref{appliouville} and discuss the resulting fusion
coefficients, which are:
\begin{eqnarray}
\label{fusionliou2deg}
{\mathcal{N}_{m,n;m',n'}}^{m'',n''} &=& \left\{ \begin{array}{ll}
1 \ \ \text{if} \ \left\{
     \begin{array}{cccc}
|m-m'|+1 \le m'' \le m+m'-1 \\
m+m'+m''+1 \equiv 0 \ [2] \\
|n-n'|+1 \le n'' \le n+n'-1 \\
n+n'+n''+1 \equiv 0 \ [2]
\end{array} \right. \\
0 \ \ \text{else}
\end{array} \right.
\end{eqnarray}
The upshot is then that we can rewrite the result for the partition
function of boundary operators in terms of the Liouville fusion
coefficients:
\begin{eqnarray}
\label{zmn2}
Z_{m,n ; m'n'} (\tau) &=& \sum_{m'',n'' \in \mathbb{N}^*}
{\mathcal{N}_{m,n ; m',n'}}^{m'',n''} \chi_{m'',n''} (-1/\tau) \\
&=& \int_0^{\infty} \sum_{m'',n'' \in \mathbb{N}^*}
{\mathcal{N}_{m,n ; m',n'}}^{m'',n''} {S_{m'',n''}}^{s} \chi_s (\tau)
ds \nonumber
\end{eqnarray}
We conclude that (since equations (\ref{zmn}) and (\ref{zmn2}) are
valid $\forall \tau \in \mathbb{C}_+$) we have $\forall s \in
\mathbb{R}_+$:
\begin{eqnarray}
\label{repring}
\frac{{S_{m,n}}^s {S_{m',n'}}^{s}}{{S_{1,1}}^s} &=& 
\sum_{m'',n'' \in \mathbb{N}^*} {\mathcal{N}_{n,m;n',m'}}^{m'',n''}
{S_{m'',n''}}^s.
\end{eqnarray}
This equation, as an equation relating modules S-matrices and fusion
coefficients could have been derived without reference to any boundary
conformal field theory. However, we will see below that its interpretation
(given above) in terms of boundary states is natural in view of a Cardy type analysis
of consistent boundary states in non-rational conformal field theory.

From equation (\ref{repring}) we see that
the modular S-matrices ${S_{m,n}}^s/{S_{1,1}}^s$ represent the
fusion ring.  From the above formula, in the unitary rational case,
one inverts the S-matrix to find the Verlinde formula. There is no
such inverse here. However, we recall that we had a similar issue when
inverting the modular transformation rules in the previous subsection
(see equation $(\ref{inversion})$). We had a complicated integral
operator, which was equal to an identity operator, and we could prove
this by analytically continuing the formula in the Liouville
momentum. We will proceed by analogy in this case. We define the
Fourier (modular) transform of the combination of modular S-matrices
in the left-hand side of equation $(\ref{repring})$ with respect to
the free momentum index $s$. (See also equation $(\ref{inversion})$.)
We then show that this transform encodes the fusion coefficients as
the residues of its poles. As before, this procedure is quite robust,
and the kernel used is the modular S-matrix in the continuous sector of
the non-rational conformal field theory. 

Let us define on the complex plane
a function that is a natural extension of the usual combination of
S-matrices used in the Verlinde formula:
\begin{eqnarray}
f(z) &=& \int_0^{\infty} \frac{{S_{m,n}}^s
  {S_{m',n'}}^{s}}{{S_{1,1}}^s} e^{- \imath \pi \sqrt{2} z s} ds \\
& & \hspace{.3cm} \text{for} \hspace{.3cm} \Im z < -\frac{1}{\sqrt{2}}
  \text{max} \left( \frac{m''}{b} + n'' b \ | \
  {\mathcal{N}_{n,m;n',m'}}^{m'',n''} \neq 0 \right) \nonumber
\end{eqnarray}
This function can be extended by analytic continuation to all the
complex plane, except for some poles. The set of poles is precisely
$\{ \pm \frac{m''}{b} \pm n'' b \ | \ m'', n'' \in \mathbb{N}^* \ , \
{\mathcal{N}_{n,m;n',m'}}^{m'',n''}~\neq~0~\}$. 
The $\pm$ signs here
are artefacts that arise due to reflection. 
We are only interested in the
$+$ signs. The fusion coefficients are given by the residues of the
function $f$:
\begin{eqnarray}
\label{verlmn}
2 \imath \pi \ \text{Residue}_{z=\left( \frac{m''}{b}+n'' b
  \right)/\imath \sqrt{2}} (f) = {\mathcal{N}_{n,m;n',m'}}^{m'',n''}
\end{eqnarray}
The function $f$, which is a natural analytic continuation of the
usual sum of modular S-matrices appearing in the Verlinde formula,
depends on a continuous parameter $z$ taking values in the complex
plane. It has poles at the values of $z$ corresponding to the
degenerate representations which occur in the fusion product $(m,n)
\otimes (m',n')$, i.e the residues coincide with the fusion
coefficients. This is a close analogue of the usual Verlinde
formula. We will check for similar properties in other cases, and in
different theories in the following.

\subsection*{Degenerate and non-degenerate representations}
Another fusion relation can be deduced from
\cite{Zamolodchikov:2001ah}, if instead of considering two degenerate
representations we consider a degenerate and a non-degenerate
representation. Indeed, the wave function for an FZZT brane,
associated to a non-degenerate representation, is:
\begin{eqnarray}
\label{psis}
\psi_s (s') = \psi_{1,1} (s') \frac{\cos (\pi s s')}{2 \sinh \left(
    \pi \frac{s'}{b} \right) \sinh \left( \pi b s' \right)} \ ,
    \hspace{1cm}  s' \neq 0
\end{eqnarray}
It satisfies $\psi_s (s') = \mathcal{R}_L (s') \psi_s (-s')$ and is
related to the modular $S$-matrix by:
\begin{eqnarray}
\label{wavefct2}
\psi_s (s') \sqrt{\mathcal{R}_L (-s')} =
\frac{{S_s}^{s'}}{\sqrt{{S_{1,1}}^{s'}}}
\end{eqnarray}
where we note once more the presence of the reflection
coefficient. Hence the partition function for a ZZ-brane and an
FZZT-brane is:
\begin{eqnarray}
Z_{m,n;s} (\tau) &=& \int_0^{\infty} \psi_{m,n} (s') \psi_s (-s')
\chi_{s'} (\tau) ds' = \int_0^{\infty} \frac{{S_{m,n}}^{s'}
  {S_s}^{s'}}{{S_{1,1}}^{s'}} \chi_{s'} (\tau) ds' \\
&=& \sum_{k=1-m,2}^{m-1} \ \sum_{l=1-n,2}^{n-1}
\chi_{s+\imath(k/b+lb)} (-1/\tau) \nonumber
\end{eqnarray}
where $\sum_{k=1-m,2}^{m-1}$ means that $k+m-1$ is always even, and
$\chi_{\tilde{s}=s+\imath(k/b+lb)}$ is the character of a
non-degenerate and non-unitary representation of complex Liouville
momentum $\tilde{s}$, given by $(\ref{chis})$. We have used the
formula\footnote{Note the similarity with $SU(2)$ character formulas
  and their use in proving the Verlinde formula.}:
\begin{eqnarray}
\frac{\sinh (\pi n b s)}{\sinh (\pi b s)} =\sum_{l=1-n,2}^{n-1}
e^{\pi l b s} 
\end{eqnarray}
The fusion coefficients calculated in Appendix~\ref{appliouville} are:
\begin{eqnarray}
\label{fuslioundeg}
{\mathcal{N}_{m,n;s}}^{s',m',n'} = \left\{ \begin{array}{ll}
1 \ \ \text{if} \ \left\{
     \begin{array}{ccccc}
1-m \le m' \le m-1 \ , \ \ m+m'+1 \equiv 0 \ [2] \\
1-n \le n' \le n-1 \ , \ \ n+n'+1 \equiv 0 \ [2] \\
s'= s \end{array} \right.
\\
0 \ \ \text{else}
\end{array} \right.
\end{eqnarray}
They agree with the computation of the partition function, in the
sense that:
\begin{eqnarray}
Z_{m,n;s} (\tau) &=& \sum_{m',n' \in \mathbb{Z}} \int_0^{\infty} ds'
\ {\mathcal{N}_{m,n;s}}^{s',m',n'} \chi_{s'+\imath(m'/b+n'b)} (-1/\tau) 
\end{eqnarray}
We conclude that $\forall s'' \in \mathbb{R}_+$:
\begin{eqnarray}
\label{vers}
\frac{{S_{m,n}}^{s''} {S_s}^{s''}}{{S_{1,1}}^{s''}} =
\sum_{m',n' \in \mathbb{Z}} \int_0^{\infty} ds'
{\mathcal{N}_{n,m;s}}^{s',m',n'} {S_{s',m',n'}}^{s''}
\end{eqnarray}
Again, it is not possible to invert this formula in the way it is done
for rational cases, but one can perform the same analysis that we
applied in the case of degenerate representations. Let us define on
the complex plane a function which is a natural analytic continuation
of the usual combination of S-matrices in the Verlinde formula:
\begin{eqnarray}
g(z) &=& \int_0^{\infty} \frac{{S_{m,n}}^{s'}
  {S_s}^{s'}}{{S_{1,1}}^{s'}} e^{-\imath \pi \sqrt{2} z s'} ds' \\
& & \hspace{.3cm} \text{for} \hspace{.3cm} \Im z < -\frac{1}{\sqrt{2}}
  \text{max} \left( \frac{m'}{b} + n' b \ | \
  {\mathcal{N}_{n,m;s}}^{s,m',n'} \neq 0 \right) \nonumber
\end{eqnarray}
The function $g$ is defined by analytic continuation when the above
integral is ill-defined. This function has poles corresponding to
representations that appear in the fusion of the representations $m,n$
and $s$, i.e verifies:
\begin{eqnarray}
\label{verlmns}
2 \imath \pi \ \text{Residue}_{z= \left( s'+ \imath \left( \frac{m'}{b}+n' b
  \right) \right)/\sqrt{2}} (g) = {\mathcal{N}_{n,m;s}}^{s',m',n'}
\end{eqnarray}
This is analogous to what we obtained in the case of degenerate
representations. Once more, a function, that is a natural analytic
continuation of the usual combination of S-matrices appearing in the
Verlinde formula in the case of rational conformal field theory,
exhibits poles precisely at representations which occur in the fusion
product $(m,n) \otimes (s)$.

\subsection*{Generalization}
Finally, let us note that the above results obtained for a
non-degenerate representation $s$ can be formally generalized to any
non-degenerate representation labeled by $\tilde{s} \in \mathbb{C}-\{
\imath \left( m/b+nb \right) \ | \ m,n \in \mathbb{Z} \}$, which
reduces to the usual non-degenerate unitary representations for $\Im
\tilde{s} =0$. The character, modular transformation and wave function
are the same as (\ref{chis}), (\ref{smss}) and (\ref{psis})
respectively, simply replacing $s$ by $\tilde{s}$. One can then show
that one obtains the non-unitary representations:
\begin{eqnarray}
Z_{m,n;\tilde{s}'} (\tau) = \sum_{p=1-m,2}^{m-1}
\sum_{q=1-n,2}^{n-1} \chi_{\tilde{s}'+\imath(p/b+qb)} (-1/\tau)
\end{eqnarray}
where the following formula has been useful:
\begin{eqnarray}
\sinh \left( \pi n b s \right) \cosh \left( \pi n' b s \right) &=&
\sum_{l=0}^{n-1} \sinh \left( \pi b s \right) \cosh \left( \pi
  (n+n'-2l-1) b s \right)
\end{eqnarray}

This expression, again, agrees with the Liouville fusion coefficients:
\begin{eqnarray}
{\mathcal{N}_{m,n ; \tilde{s}'}}^{\tilde{s}''} = \left\{ \begin{array}{ll}
1 \ \ \text{if} \ \left\{
     \begin{array}{cccc}
\Im (\tilde{s}''-\tilde{s}') = p/b + qb \ \text{with} \ p,q \in
\mathbb{Z} \\
1-m \le p \le m-1 \ \text{and} \ p+m+1 \equiv 0 \ [2] \\
1-n \le q \le n-1 \ \text{and} \ q+n+1 \equiv 0 \ [2] \\
\Re \tilde{s}'' = \Re \tilde{s}'
\end{array} \right. \\
0 \ \ \text{else}
\end{array} \right.
\end{eqnarray}
Formulas similar to equations (\ref{vers}) and (\ref{verlmns}) can
also be obtained. In deriving these formulas we briefly explored the
idea of extending the Verlinde formula into the domain of complexified
momenta. We will comment further on this possibility later on.

In summary, we did obtain a generalization of the Verlinde formula,
applicable to the fusion of degenerate representations with generic
ones. The formula requires an analytic continuation in the Fourier
transformed free index of the formula that shows that the modular
S-matrices give a representation of the fusion coefficients. The
fusion coefficients then appear as non-trivial residues of poles in
the transformed function on the complex plane.  We will see how this
pattern persists in other non-rational conformal field theories.

Our analysis shows that the relation between modular S-matrices and
fusion coefficients is in fact more general than the relations encoded
in  the standard boundary states.


\subsection{The hyperbolic three-plane $H_3^+$}
For our next example, we turn to the hyperbolic three-plane, and
summarize the brane computation for a "spherical" brane
\cite{Ponsot:2001gt} for starters, with finite representations in the
open string channel. We will see that there are strong similarities
with the Liouville case. We then go on to generalize the analysis to
other representations.

Several results concerning the $H_3^+$ theory, including fusion, are
collected in Appendix~\ref{apph3}.

We will consider continuous non-degenerate representations of
$SL(2,\mathbb{R})$, labeled by $j = \frac{1}{2} + \imath \lambda$
where $\lambda \in \mathbb{R}_+$, and finite degenerate
representations (generated by the current algebra from a ground state
with a $2J+1$-fold degeneracy), labeled by $s = \pi b^2 (2J+1)$ where
$2J$ is an integer. The degenerate representation $J=0$ will play the
role of the identity in the Verlinde formula. The characters of a
non-degenerate and of a degenerate representation are respectively:
\begin{eqnarray}
\chi_{\lambda} (\tau) = \frac{q^{b^2
    \lambda^2}}{\eta(\tau)^3} \ , \hspace{1cm} \chi_J (\tau) = (2J+1)
    \frac{q^{-b^2 (2J+1)^2/4}}{\eta(\tau)^3}
\end{eqnarray}
where $b^2 = 1/k$ and the level $k$ is real and strictly positive.

The modular transformations of these characters are (see
\cite{Giveon:2001uq} for the degenerate case):
\begin{eqnarray}
\chi_J \left( - \frac{1}{\tau} \right) &=& \int_0^{\infty}
{S_J}^{\lambda} \chi_{\lambda} (\tau) d\lambda \ , \hspace{1cm}
{S_J}^{\lambda} = 4 \sqrt{2} b \lambda \sinh(2 \pi b^2 \lambda (2J+1)) \\
\chi_{\lambda} \left( -\frac{1}{\tau} \right) &=& \int_0^{\infty}
  {S_{\lambda}}^{\lambda'} \chi_{\lambda'} (\tau) d \lambda' \ , \hspace{1cm}
  {S_{\lambda}}^{\lambda'} = - \frac{2 \sqrt{2} b}{\imath \tau} \cos
  (4 \pi b^2 \lambda \lambda')
\end{eqnarray}
Note that ${S_{\lambda}}^{\lambda'}$ depends on $\tau$. Notice also
the similarity between the $H_3^+$ modular transformations and the
Liouville ones given in (\ref{smss}) and (\ref{degmod}).

\subsection*{Degenerate representations}
For a ''spherical brane'', the one-point function is:
\begin{eqnarray}
\langle \Phi^j (x|z) \rangle_s = - \frac{(1+x
    \bar{x})^{2j-2}}{|z-\bar{z}|^{2 \Delta_j}} \Gamma (1+(2j-1) b^2)
    \frac{\sin s (2j-1)}{\sin s} \frac{\nu_b^j}{2 \pi \Gamma (1-b^2)}
\end{eqnarray}
with $\nu_b = \frac{\Gamma (1-b^2)}{\Gamma (1+b^2)}$. This
wave-function satisfies the usual reflection property:
\begin{eqnarray}
\langle \Phi^j (x|z) \rangle_s = \frac{2j-1}{\pi} \mathcal{R}(j)
\int_{\mathbb{C}} d^2 y |x-y|^{4j-4} \langle \Phi^{1-j} (y|z) \rangle_s
\end{eqnarray}
where $\mathcal{R}(j)$ is\footnote{Our notation here is different from
the one in Appendix \ref{apph3} in order to have the normalization
\newline
$\mathcal{R}(j) \mathcal{R}(1-j)=1$}:
\begin{eqnarray}
\mathcal{R}(j) &=& - \frac{\Gamma (1+(2j-1) b^2)}{\Gamma
  (1-(2j-1) b^2)} \ \nu_b^{2j-1}
\end{eqnarray}
We introduce the boundary state\footnote{We choose a normalization
  different from \cite{Ponsot:2001gt}, so that the partition function
  is normalized with respect to the fusion of representations.}:
\begin{eqnarray}
_B\langle s | j;x \rangle = 2 \sin s \sqrt{\frac{2 \sqrt{2} b \pi}{\sin \pi b^2}}
{_B\langle} \Phi^j \left( x|\frac{\imath}{2} \right) \rangle_s
\end{eqnarray}
Note that $\left( _B\langle s | j;x \rangle \right)^* = \langle j;x |
s \rangle_B =_B\langle s | 1-j;x \rangle$.

The boundary state is related to the modular transformation in the
following way:
\begin{eqnarray}
\label{wavefct3}
_B\langle s | j;x \rangle \sqrt{\mathcal{R}(1-j)} =
\frac{{S_J}^{\lambda}}{\sqrt{{S_0}^{\lambda}}} \frac{(1 + x
  \bar{x})^{2j-2}}{\sqrt{\pi}}
\end{eqnarray}
The annulus amplitude for two ''spherical branes'' is then:
\begin{eqnarray}
_B\langle s' | q'^{H/2} | s \rangle_B &=& \int_{\mathbb{S}} dj
\int_{\mathbb{C}} d^2x \ _B\langle s' | j;x
\rangle \langle j;x | s \rangle_B \ \chi_j (q') \\
&=& \int_0^{\infty} d \lambda \frac{{S_J}^{\lambda}
  {S_{J'}}^{\lambda}}{{S_0}^{\lambda}} \chi_{\lambda} (q') =
\sum_{J''=|J-J'|}^{J+J'} \ \chi_{J''} (q)
\end{eqnarray}
where the momenta are given by $s = \pi b^2 (2J+1)$, $s' = \pi b^2
(2J'+1)$, and $H = \frac{L_0 + \bar{L}_0}{2} - \frac{c}{24}$ is the
Hamiltonian on the plane and $\mathbb{S} = \{ \frac{1}{2}+\imath
\lambda | \lambda \in \mathbb{R}_+ \}$ so that $\int_{\mathbb{S}} dj =
\int_0^{\infty} d \lambda$.

When studying Liouville theory, we noticed that it was possible to
rewrite the partition function of boundary operators in terms of the
fusion coefficients. This property is shared by the $H_3^+$
theory. Indeed, the fusion coefficients for finite degenerate
representations found in Appendix~\ref{apph3} are:
\begin{eqnarray}
\label{fusdegh3}
{\mathcal{N}_{u;u'}}^{u''} = \left\{ \begin{array}{ll}
1 \ \ \text{if} \ \left\{
     \begin{array}{ccccc}
|u-u'|+1 \le u'' \le u+u'-1 \\
u+u'+u''+1 \equiv 0 \ [2] \\
\end{array} \right.
\\
0 \ \ \text{else}
\end{array} \right.
\end{eqnarray}
which can be rewritten in the following way, if we note $u=2J+1$,
$u'=2J'+1$ and $u''=2J''+1$:
\begin{eqnarray}
{\mathcal{N}_{J;J'}}^{J''} = \left\{ \begin{array}{ll}
1 \ \ \text{if} \ \left\{
     \begin{array}{ccccc}
|J-J'| \le J'' \le J+J' \\
J+J'+J'' \in \mathbb{N} \\
\end{array} \right.
\\
0 \ \ \text{else}
\end{array} \right.
\end{eqnarray}
Hence:
\begin{eqnarray}
_B\langle s' | q'^{H_P/2} | s \rangle_B =\sum_{J'' \in \frac{1}{2}
  \mathbb{N}} {\mathcal{N}_{J;J'}}^{J''} \chi_{J''} (q)
\end{eqnarray}
Just like in the Liouville case, it is possible to construct a natural
analytic continuation of the usual sum of modular S-matrices appearing
in the Verlinde formula, whose poles correspond to representations
found in the fusion of $s$ and $s'$ (see equations $(\ref{verlmn})$
and $(\ref{verlmns})$). Calculations would closely follow the lines of
the Liouville case, hence we do not reproduce them here.

Nevertheless, it is important to remark that although degenerate
representations of $SL(2,\mathbb{R})$ are related, via Hamiltonian
reduction, to degenerate representations of the Virasoro algebra, this
is not the case precisely for the finite representations of
$SL(2,\mathbb{R})$ \cite{Kato:1991zz}, and therefore the above check
of the Verlinde formula in the case of finite representations does
represent independent evidence for its validity.

\subsection*{Degenerate and non-degenerate representations}

We will now consider, in analogy of the Liouville case, the annulus
amplitude between a degenerate and a non-degenerate
representation. For this purpose, we recall the one-point function for
a continuous  $AdS_2$ brane \cite{Ponsot:2001gt}:
\begin{eqnarray}
\langle \Phi^j (x|z) \rangle_r = \frac{| x+\bar{x} |^{2j-2}}{| z-\bar{z} |^{2
  \Delta_j}} \frac{A_b \nu_b^{j-1/2}}{\pi b^2} \Gamma (1+(2j-1) b^2)
  e^{- r (2j-1) \sigma}
\end{eqnarray}
where $\sigma = \text{sign} (x+\bar{x})$ and $2 \sqrt{2} |A_b|^2 =
\pi^2 b^3$. This one-point function is related to a non-degenerate
representation $j = \frac{1}{2} + \imath R$ where $r=2 \pi b^2 R$.
The boundary state is defined as\footnote{We choose a normalization
  different from \cite{Ponsot:2001gt}, so that the partition function
  is normalized with respect to the fusion of representations.}:
\begin{eqnarray}
_B\langle r | j;x \rangle = \frac{2^{-1/4} b^{3/2}}{A_b}
{_B\langle} \Phi^j \left( x|\frac{\imath}{2} \right) \rangle_r
\end{eqnarray}
As was pointed out in \cite{Ponsot:2001gt}, it is necessary to
define a regularized boundary state in order to be able to calculate
the annulus amplitude $_{B, reg.}\langle r' | q^{H_p /2} | r
\rangle_{B,reg.}$. However, this regularization is not needed for the
calculation of $_B\langle r | q'^{H_p /2} | s \rangle_B$. (It can be
checked that the result would be the same using the regularized state
$_{B, reg.}\langle r |$ and then taking the well-defined limits in all
the cut-offs.)

In the following, we will Fourier-transform the boundary states like
in \cite{Ponsot:2001gt}, because calculations are simpler in this
basis and also because it is the one that should be used for
regularizing the $_B\langle r |$ boundary state. Therefore:
\begin{eqnarray}
_B\langle r | j;n,p \rangle &=& \int_{\mathbb{C}} d^2x e^{-\imath n
  \text{arg} (x)} |x|^{-2j-\imath p} {_B\langle} r | j;x \rangle = 2 \pi
  \delta (p) A(j,n|r)
\end{eqnarray}
where $n \in \mathbb{Z}$, $p \in \mathbb{R}$ and:
\begin{eqnarray}
A(j,n|r) &=& 2^{3/4} b^{-1/2} \nu_b^{j-\frac{1}{2}} \Gamma(1+(2j-1)
  b^2) \frac{\Gamma (2j-1)}{\Gamma (j+\frac{n}{2}) \Gamma
  (j-\frac{n}{2})} \\
& & \times \left( \frac{1+(-1)^n}{2} \cos (2 \lambda r) -
  \frac{1-(-1)^{n}}{2} \imath \sin(2 \lambda r) \right) \nonumber
\end{eqnarray}

The Fourier transform of a continuous $s$ boundary state will also be
useful:
\begin{eqnarray}
_B\langle s | j;n,p \rangle &=& - 2 \sqrt{\frac{2 \sqrt{2}}{b}}
\nu_b^{j-\frac{1}{2}} \Gamma (1+(2j-1) b^2) \sin s (2j-1) \\
& & \times \frac{\Gamma
  (1-j-\frac{\imath p}{2}) \Gamma (1-j+\frac{\imath p}{2})}{\Gamma
  (2-2j)} \delta_{n,0} \nonumber
\end{eqnarray}

We then calculate the annulus amplitude for a ''spherical'' brane and
an $AdS_2$ brane:
\begin{eqnarray}
_B\langle r | q'^{H_P /2} | s \rangle_B &=& - \int_{\mathbb{S}} dj
\frac{1}{(2 \pi)^2} \sum_{n \in \mathbb{N}} \int_{\mathbb{R}} dp
\ _B\langle r | j;n,p \rangle \langle j;n,p | s
\rangle_B \ \chi_j (q')\\
&=& \imath \tau \int_0^{\infty} \frac{{S_R}^{\lambda}
  {S_J}^{\lambda}}{{S_{J=0}}^{\lambda}} \ \chi_{\lambda} (q') d \lambda
= \imath \tau \sum_{J'+\frac{1}{2}=-J}^{J} \chi_{R+\imath (2J'+1)/2} (q)
\label{partitionrs}
\end{eqnarray}
where $\chi_{R+\imath (2J'+1)/2} (q) = \frac{q^{b^2 (R+\imath
  (2J'+1)/2)^2}}{\eta (\tau)^3}$.
Once again, it is possible to rewrite this annulus amplitude in terms
of the fusion coefficients, which for a degenerate and a
non-degenerate representations where found in Appendix~\ref{apph3} to
be:
\begin{eqnarray}
\label{fusionnondegh3}
{\mathcal{N}_{R;J}}^{R',J'} = \left\{ \begin{array}{ll}
1 \ \ \text{if} \ \left\{
     \begin{array}{ccccc}
-J \le J'+\frac{1}{2} \le J \ , \hspace{.5cm} J+J'+\frac{1}{2} \in
 \mathbb{N} \\
R=R'
\end{array} \right.
\\
0 \ \ \text{else}
\end{array} \right.
\end{eqnarray}
The annulus amplitude is then:
\begin{eqnarray}
_B\langle r | q'^{H_P /2} | s \rangle_B = \imath \tau
  \int_0^{\infty} dR' \sum_{J' \in \frac{1}{2} \mathbb{N}}
  {\mathcal{N}_{R;J}}^{R',J'} \chi_{R'+\imath (2J'+1)/2} (q)
\end{eqnarray}

Finally, in analogy with the Liouville case, it is straightforward to
generalize the above results for non-degenerate representations $R$ to
non-degenerate representations $\tilde{R} = R+\imath (2J+1)/2$.

We conclude that we find in the $H_3^+$ conformal field theory the
same Verlinde type relations between the S-matrices and the
fusion coefficients as in the bosonic Liouville theory for both
degenerate and non-degenerate representations.


\subsection{The supersymmetric coset $SL(2,\mathbb{R})/U(1)$}

In the case of the supersymmetric coset $SL(2,\mathbb{R})/U(1)$,
 we summarize and extend the brane calculations made for
the finite representations in \cite{Eguchi:2003ik,Israel:2004jt}. We
also compute the overlap between branes associated to finite
representations and branes associated to continuous representations.

Our notations are as follows: for continuous representations we have
$j = \frac{1}{2} + \imath P$ with $P \in \mathbb{R}_+$, for discrete
representations $2j \in \mathbb{N}$, $0 < j < \frac{k+1}{2}$, and for
finite representations $j = \frac{1-u}{2}$ with $u \in
\mathbb{N}^*$. To avoid confusion, we will label continuous, discrete
and finite representations by respectively $P$, $j$ and $u$, unless
otherwise stated. The characters of these representations can be found
for instance in \cite{Israel:2004jt}. The central charge of the
supersymmetric coset is $c = 3 + \frac{6}{k}$, where the level $k$ is
assumed to be strictly positive, real and non-rational.

The modular S-matrix for characters of degenerate representations
is given by:
\begin{eqnarray}
{S_u}^{P,w} &=& (-1)^{w(u-1)} \frac{2 \sinh \left( 2 \pi P \right) \sinh
  \left( 2 \pi \frac{P}{k} u \right)}{\cosh \left( 2 \pi P \right) +
  \cos \left( \pi k w \right)} \nonumber \\
{S_u}^{j,w} &=& (-1)^{w(u-1)} 2 \sin \left( \frac{\pi}{k} (2j-1) u
  \right)
\end{eqnarray}
The character for a continuous representation is:
\begin{eqnarray}
ch_c^{\text{NS}} \left( P,\frac{w k}{2};\tau \right) =
q^{\frac{P^2}{k}+\frac{w^2 k}{4}} \ \frac{\theta_3(\tau)}{\eta(\tau)^3}
\end{eqnarray}
Its modular transformation is:
\begin{eqnarray}
\label{transfocont}
\sum_{n \in \mathbb{Z}} ch_c^{\text{NS}} \left( P,\frac{w k}{2}+n;
  -\frac{1}{\tau} \right) = \sum_{w' \in \mathbb{Z}} \int_0^{\infty}
dP' \ 2 \ e^{- \imath \pi w w' k} \cos \left( \frac{4 \pi P P'}{k} \right)
ch_c^{\text{NS}} \left( P',\frac{w' k}{2};\tau \right) \nonumber \\
\end{eqnarray}
Note that in the following we put $z = e^{2 \imath \pi \nu} = 1$.

\subsection*{Degenerate representations}
The one-point function for a D0-brane is:
\begin{eqnarray}
\psi_u^{\text{NS}} (j,w) = \psi_1 (j,w) (-1)^{w(u-1)} \frac{\sin \left(
    \frac{\pi}{k} (2j-1) u \right)}{\sin \left( \frac{\pi}{k} (2j-1)
    \right)}
\end{eqnarray}
where we have used the expression for the one-point function of the
$u=1$ brane:
\begin{eqnarray}
\psi_1^{\text{NS}} (j,w) = \frac{(-1)^w}{\sqrt{k}} \nu^{\frac{1}{2}-j}
    \frac{\Gamma(j+\frac{kw}{2}) \Gamma(j-\frac{kw}{2})}{\Gamma(2j-1)
    \Gamma(1+\frac{2j-1}{k})}
\end{eqnarray}
where $j$ may represent either continuous or discrete representations,
and $\nu = \frac{\Gamma \left( 1-\frac{1}{k} \right)}{\Gamma \left(
    1+\frac{1}{k} \right)}$.

From this wave function we define a reflection amplitude:
\begin{eqnarray}
\mathcal{R}^{\text{NS}} (j,w) = \nu^{1-2j} \frac{\Gamma(1-2j)
  \Gamma(1+\frac{1-2j}{k})}{\Gamma(2j-1) \Gamma(1+\frac{2j-1}{k})}
  \frac{\Gamma(j+\frac{kw}{2})
  \Gamma(j-\frac{kw}{2})}{\Gamma(1-j+\frac{kw}{2})
  \Gamma(1-j-\frac{kw}{2})} \nonumber
\end{eqnarray}
which satisfies $\mathcal{R}^{\text{NS}} (j,w) \mathcal{R}^{\text{NS}}
(1-j,-w)$ and $\psi_{u} (j) = \mathcal{R} (j,w) \psi_u (1-j)$. Note
that for a continuous representation, $\mathcal{R}^{\text{NS}} (P,w)$
is just a phase.

The wave function and the modular S-matrix are then related. For a
continuous representation we have:
\begin{eqnarray}
\label{wavefct4}
\psi_u^{\text{NS}} (P,w) \sqrt{\mathcal{R}^{\text{NS}} (-P,-w)} &=&
\frac{{S_u}^{P,w}}{\sqrt{{S_1}^{P,w}}}
\end{eqnarray}
For a discrete representation, things are a little more complicated,
because an infinity appears in the wave-function. Because of this, we
will not relate the modular S-matrix to the wave-function, but rather
to the residue of a product of wave functions. More precisely:
\begin{eqnarray}
2 \pi \ \text{Res} \left( \psi_u^{\text{NS}} (j,w) \psi_{u'}^{\text{NS}} (1-j,-w) \right) =
\frac{{S_u}^{j,w} {S_{u'}}^{j,w}}{{S_1}^{j,w}}
\end{eqnarray}
with $j = j_r = -r + \frac{k w_r}{2}$ and $r,w \in \mathbb{Z}$.
Note that it is not surprising to find a residue here, because the
discrete term in the modular transformation is obtained as a residue
(see \cite{Miki:1989ri,Israel:2004xj}).

The partition function for two D0-branes calculated in
\cite{Eguchi:2003ik,Israel:2004jt} can be expressed as:
\begin{align}
Z_{u,u'}^{NS} \left( -\frac{1}{\tau} \right) &= \sum_{w \in
    \mathbb{Z}} \int_0^{\infty} dP \psi_u^{\text{NS}} (-P,-w)
    \psi_{u'}^{\text{NS}} (P,w)
    ch_c^{\text{NS}} \left( P, \frac{w k}{2} ; \tau \right) \\
& \hspace{2cm} + 2 \pi \sum_{r \in
  \mathbb{Z}} \text{Res} \left( \psi_u^{\text{NS}} (1-j,-w)
    \psi_{u'}^{\text{NS}} (j,w)
\right) ch_d^{\text{NS}} (j,r; \tau) \nonumber \\
&= \sum_{w \in \mathbb{Z}} \int_0^{\infty} dP
\frac{{S_u}^{P,w} {S_{u'}}^{P,w}}{{S_1}^{P,w}} ch_c^{\text{NS}} \left(
  P, \frac{w k}{2} ; \tau \right) + \sum_{r \in \mathbb{Z}}
\frac{{S_u}^{j,w} {S_{u'}}^{j,w}}{{S_1}^{j,w}} ch_d^{\text{NS}} (j,r;
\tau) \nonumber \\
&= \sum_{u''=|u-u'|+1}^{u+u'-1} Z_{u'',1} \left(- \frac{1}{\tau}
    \right)
\end{align}
Modular transforming the last line and identifying the continuous and
discrete contributions, we find that:
\begin{eqnarray}
\frac{{S_u}^{P,w} {S_{u'}}^{P,w}}{{S_1}^{P,w}} &=& \sum_{u'' \in
  \mathbb{Z}} {\mathcal{N}_{u,u'}}^{u''} {S_{u''}}^{P,w} \ ,
  \hspace{1cm} \forall P \in \mathbb{R}_+ \ , \ w \in \mathbb{Z} \\
\frac{{S_u}^{j,w} {S_{u'}}^{j,w}}{{S_1}^{j,w}} &=& \sum_{u'' \in
  \mathbb{Z}} {\mathcal{N}_{u,u'}}^{u''} {S_{u''}}^{j,w} \ ,
\hspace{1cm} \forall j \in \frac{1}{2} \mathbb{Z} \ , \ w \in
\mathbb{Z}
\end{eqnarray}
where we used the fusion coefficients in equation $(\ref{fusdegh3})$.

In analogy with the Liouville and the $H_3^+$ case (note that we keep
using the same kernel, wich is just the modular transformation in the
continuous sector), we define a function of a complex variable $z$
that is an analytic continuation of the usual combination of
S-matrices in the Verlinde formula, i.e:
\begin{eqnarray}
f(z) = \int_0^{\infty} \frac{{S_u}^{P,w} {S_{u'}}^{P,w}}{{S_1}^{P,w}}
(-1)^w e^{-2 \imath \pi z (P+\imath k w/2)} dP
\end{eqnarray}
The function $f$ can be analytically continued to the whole complex
plane except for some poles at $z = \pm \imath \frac{u''}{k} + \imath
l$ with $l \in \mathbb{N}^*$ and $u''$ in the fusion of $u$ and
$u'$. They correspond to representations $u \pm l k$ which are
presumably spectral-flowed representations.
The fusion coefficients are once more given by residues of the
function $f$:
\begin{eqnarray}
2 \imath \pi \ \text{Residue}_{z=-\imath u''/k} (f) =
{\mathcal{N}_{u,u'}}^{u''}
\end{eqnarray}
where ${\mathcal{N}_{u,u'}}^{u''}$ was given in $(\ref{fusdegh3})$.

The case of discrete representations is easier to handle. Indeed, we
can use the following useful relation:
\begin{eqnarray}
\frac{1}{2} \sum_{r \in \mathbb{Z}} {S_u}^{j_r,w_r} {S_{u'}}^{j_r,w_r} &=& \sum_{l
\in \mathbb{Z}} (-1)^l \left( \delta \left( l-\frac{u-u'}{k} \right)
- \delta \left( l+\frac{u+u'}{k} \right) \right) \nonumber \\
&=& \left\{ \begin{array}{ll}
\delta (0) \ \ \text{if} \ u=u' \\
0 \ \ \text{else}
\end{array} \right.
\end{eqnarray}
from which we deduce the following Verlinde type formula:
\begin{eqnarray}
\sum_{r \in \mathbb{Z}} \frac{{S_u}^{j_r,w_r} {S_{u'}}^{j_r,w_r}
  \frac{1}{2} {S_{u''}}^{j_r,w_r}}{{S_1}^{j_r,w_r}} = \delta (0) \times
  {\mathcal{N}_{u,u'}}^{u''}
\end{eqnarray}
This looks very much like a Verlinde formula in rational cases, except
for the appearance of the infinite factor $\delta(0)$. This factor is
due to us neglecting the $U(1)$ quantum number as indicated below
equation (\ref{transfocont}).

\subsection*{Degenerate and non-degenerate representations}

The wave-function of an A-brane as given in \cite{Hosomichi:2004ph}
is\footnote{We use a slightly different normalization in order to
  preserve the form of equation $(\ref{wavecos})$.}:
\begin{eqnarray}
\psi_J (j,w) = \frac{\pi}{\sqrt{k}} \nu^{\frac{1}{2}-j}
\frac{\Gamma (1-2j) \Gamma \left( 1-\frac{2j-1}{k} \right)}{\Gamma
  \left( 1-j+\frac{wk}{2} \right) \Gamma \left( 1-j-\frac{wk}{2}
  \right)} \cos \left( \frac{\pi}{k} (2j-1) (2J-1) \right)
\end{eqnarray}
The modular S-matrix elements are:
\begin{eqnarray}
{S_J}^{j,w} = 2 \cos \left( \frac{\pi}{k} (2j-1) (2J-1) \right)
\end{eqnarray}
And the wave-function is again related to the S-matrix in the following way:
\begin{eqnarray}
\label{wavecos}
\label{wavefct5}
\psi_J (j,w) \sqrt{\mathcal{R}(1-j,w)} =
\frac{{S_J}^{j,w}}{\sqrt{{S_1}^{j,w}}}
\end{eqnarray}
In the following, we will use the notation $2J-1 = 2 \imath P$, while $2j-1 =
2 \imath P'$.

The partition function is:
\begin{eqnarray}
Z^{NS}_{u;P} \left( -\frac{1}{\tau} \right) &=& \sum_{w \in
    \mathbb{Z}} \int_0^{\infty} dP' \psi_u (-P',-w) \psi_P (P',w)
    ch_c^{\text{NS}} \left( P', \frac{w k}{2} ; \tau \right) \\
&=& \sum_{w \in \mathbb{Z}} \int_0^{\infty} dP' \frac{{S_u}^{P',w}
  {S_P}^{P',w}}{{S_1}^{P',w}} ch_c^{\text{NS}} \left( P', \frac{w k}{2} ;
  \tau \right) \nonumber \\
&=& \sum_{u'=1-u,2}^{u-1} \ \sum_{n \in \mathbb{Z}} ch_c^{\text{NS}}
    \left( P+\imath \frac{u}{2},\frac{u-1}{2}+n,-\frac{1}{\tau} \right) \nonumber
\end{eqnarray}
We may once more define a function that is a generalization of the
usual combination of S-matrices appearing in the Verlinde formula for
rational cases:
\begin{eqnarray}
g(z) = \int_0^{\infty} \frac{{S_u}^{P',w} {S_P}^{P',w}}{{S_1}^{P',w}}
e^{-2 \imath \pi z P'} dP'
\end{eqnarray}
This function is defined through analytic continuation when the above
integral is ill-defined. The function $g$ is then a meromorphic
function which has poles for $z = \frac{\pm 2P+\imath u'}{k}$, which
corresponds to representations which belong to the fusion of the
representations $u$ and $J$ (the $\pm$ sign is an artefact of the
calculation because the sign of $P$ does not actually matter). More
precisely:
\begin{eqnarray}
2 \imath \pi \text{Residue}_{z=(2P+\imath u')/k} (g) = {\mathcal{N}_{u;P}}^{P,u'}
\end{eqnarray}
where $P,u'$ corresponds to a representation for which $j =
\frac{1-u'}{2}+\imath P$, and:
\begin{eqnarray}
{\mathcal{N}_{u;P}}^{P',u'} = \left\{ \begin{array}{ll}
1 \ \ \text{if} \ \left\{
     \begin{array}{ccccc}
1-u \le u' \le u-1 \ , \ \ u+u'+1 \equiv 0 \ [2] \\
P'= P \end{array} \right.
\\
0 \ \ \text{else}
\end{array} \right.
\end{eqnarray}
Results for the supersymmetric coset have therefore proven to be very
analogous to the ones obtained for the Liouville and $H_3^+$ theories.

\subsubsection*{Bosonic coset}
Results in the case of the bosonic coset are expected to be similar to
the case of the supersymmetric coset, and can be obtained using the
calculations in 
\cite{Ribault:2003ss, Israel:2004xj, Fotopoulos:2004ut}. We will not go
through this example explicitly here.

\section{Conclusions}

In summary, we have seen that it is possible to obtain an analogue of
the Verlinde formula in the non-rational conformal field theories we
studied, namely the bosonic Liouville theory, the hyperbolic
three-plane $H_3^+$ and the superconformal $SL(2,\mathbb{R})/U(1)$ coset. The
formula we found applies only to a subset of representations,
involving the fusion (or modular transformation matrices) of what we
could generically call degenerate representations. These
representations are characterized by null vectors appearing in the
associated chiral Verma module. It is known that these representations
are crucial when deriving differential equations for the (bulk and
boundary) correlation functions of the non-degenerate fields from
postulating decoupling of null vectors. Thus, degenerate
representations have already been put to good use to determine the
structure of non-rational conformal field theories through
differential methods. One can view the results on the generalized
Verlinde formula for degenerate representations as laying bare some of
the algebraic structure underlying solutions for the unitary sector of
non-rational conformal field theories (even though degenerate
representations may not be contained in the unitary conformal field
theory spectrum).\footnote{As is well-known, non-unitary sectors of
  conformal field theories are not only interesting for
  two-dimensional physics (e.g. the Yang-Lee singularity), but also
  arise in the covariant quantization of unitary string theories. One
  needs to keep the distinction between unitary conformal field
  theories and unitary string theories (which implement physical state
  conditions)  carefully in mind.}

Moreover, we have seen one further example of a phenomenon which is
ubiquitous. Instead of concentrating on quantities which depend on a
real variable parameterizing the unitary (continuous, say) spectrum of
a non-rational conformal field theory, we consider functions of a
complexified parameter. This is familiar from the analysis of discrete
contributions to partition functions
\cite{Maldacena:2000kv,Hanany:2002ev}, from the determination of the
moduli space of FZZT branes in minimal string theories and its
properties \cite{Seiberg:2003nm}, from the determination of the fusion
of degenerate representations from the fusion of non-degenerate ones,
and now from the fact that the generalized Verlinde formula is based
on this same idea, rendering a function regular by complexifying a
momentum, and then extending the domain of definition of the
regularized function over the complex plane. We have given some new
examples of connections of the Verlinde type between modular
transformation properties and fusion of non-unitary representations,
associated to complexified momenta. Although these complexifications
may seem formal on occasion, we believe they point towards quite
generic structures underlying these non-rational conformal field
theories, which may be more naturally defined in a complexified
external parameter space (e.g. bulk coupling constants, external
momenta, boundary coupling constants).

In the context of string theoretic applications of the branes of
non-rational conformal field theories, it is clear that we expect a
generalized Verlinde formula to be at work for branes that are
localized (or boundary state calculations involving at least one
localized brane). The localization of the associated open string
avoids having to deal with volume divergences (see
e.g.\cite{Ponsot:2001gt}), which is crucial to our
calculations\footnote{It would be interesting to find regularizations
  of brane partition functions that grow like the volume of a
  non-compact space, and that are consistent with all the symmetries
  of the theory.}.

In performing our calculations, we have taken the opportunity to check
the generalized Cardy formula relating the one-point function of a brane to the
reflection coefficient and modular S-matrix \cite{Teschner:2000md}
(see equations (\ref{wavefct1}), (\ref{wavefct2}), (\ref{wavefct3}),
(\ref{wavefct4}) and (\ref{wavefct5})), suggesting its general validity.
It is quite interesting that such a general formula can be written down, which
would provide localized branes for any non-rational conformal field theory.

Apart from the new modular transformation properties obtained in this paper ,  one can
apply the techniques developed here to a further variety of
non-rational conformal field theories, including theories with $N=4$
superconformal symmetry, with $N=2$ extended superconformal symmetry
at central charge $c=9$, the $H_{2n}$ models (e.g. the localized
$S(-1)$ brane in $H_4$ \cite{D'Appollonio:2004pm}), and the bosonic $SL(2,\mathbb{R})/U(1)$
model.  Further open problems include an analysis of the mechanics of
both fusion and modular transformations at rational values for the
central charge.

One hope is that an understanding of these sectors that connect
analytic to algebraic properties of non-rational conformal field
theories will allow for more efficient algebraic constructions of
boundary conformal field theories. These in turn would allow for a
better understanding of for instance D-branes in non-compact
Calabi-Yau's and the spectrum of open strings living on them, to name
only one application.

\section{Acknowledgments}
We thank Costas Bachas, Dan Israel, Elias Kiritsis and Ari Pakman for discussions on
these and related topics, and especially Mathias Gaberdiel for a
question motivating us further.

\appendix


\section{Fusion in the $H_3^+$ theory}
\label{apph3}
\subsection*{Introductory remarks}
As a brief introduction to the analysis of fusion in the bosonic
Liouville and the hyperbolic three-plane $H^+_3$ model, we recall that
fusion was first analyzed for minimal models in~\cite{Belavin:1984vu}
while for Wess-Zumino-Witten models, the fusion rules for integrable
highest weight representations were obtained in~\cite{Gepner:1986wi}.
We also note that an algebraic analysis of fusion in terms of
generalized tensor products was performed in
\cite{Gaberdiel:1993td}. A review of fusion may be found in
e.g.\cite{Fuchs:1997af}. Note that fusion has not been understood in
all cases, even in the case of rational conformal field theories. See
e.g.\cite{Gaberdiel:2001ny} for an example of fusion for
Wess-Zumino-Witten models at rational level.

For the cases we treat below, fusion is fairly
well-understood. However for bosonic Liouville our more detailed
analysis of degenerate fusion will lay bare some less widely appreciated
features. We derive the fusion relations in some detail since they are
used in the bulk of the paper to perform checks on the Verlinde
formula.

\subsection*{The fusion}
We approach the problem of fusion in $H_3^+$ directly via the
three-point function and the operator product expansion. The
three-point functions of the supersymmetric coset model can then be
reconstructed by starting from the three-point function of the
Euclidean $SL(2,\mathbb{R})$ model, where we gauge a $U(1)$ direction
to obtain an analogue of the cigar coset model, and where we can add
an other $U(1)_R$ direction to regain supersymmetry. We start out by
recapitulating the three-point function of the Euclidean
$SL(2,\mathbb{R})$ model, i.e. the $H_3^+$ model
\cite{Teschner:1997ft,Hosomichi:2001fm,Satoh:2001bi,Maldacena:2001km}.

The $H_3^+$ model is classically defined by the Lagrangian:
\begin{eqnarray}
\mathcal{L} = \frac{k+2}{\pi} \left( \partial \Phi \bar{\partial} \Phi
  + e^{2 \Phi} \partial \gamma \bar{\partial} \bar{\gamma} \right)
\end{eqnarray}
where $\Phi$, $\gamma$ and $\bar{\gamma}$ are Poincare coordinates on
Euclidean $AdS_3$.
The central charge of the model is $c=3+\frac{6}{k}$, and the classical
regime is obtained in the limit $k \rightarrow \infty$.

Primary fields are of the form:
\begin{eqnarray}
\Phi_j (x;z) = \left( e^{- \Phi} + e^{\Phi} | \gamma - x |^2 \right)^{-2j}
\end{eqnarray}
where $x$ and $\bar{x}$ are auxiliary coordinates that keep track of
the $H_3^+$ symmetry. The operator $\Phi_j (x; z)$ has conformal
weight $h_j = - \frac{j (j-1)}{k}$, where $j$ is called the spin.

We will be interested in two kinds of representations of
$SL(2,\mathbb{R})$, namely continuous (non-degenerate and unitary)
representations for which $j = \frac{1}{2} + \imath \lambda$ with
$\lambda \in \mathbb{R}_+^*$, and finite (degenerate) representations
for which $j = \frac{1-u}{2}$, with $u \in \mathbb{N}^*$ (non-unitary
unless $u=1$)

Note that $h_j$ is invariant under $j \rightarrow 1-j$, which suggests
that fields with spin $j$ and $1-j$ may be related, and indeed they
satisfy the reflection relation:
\begin{eqnarray}
\label{reflectionj}
\Phi_j (x;z) = \frac{\mathcal{R}(j)}{\pi}
\int_{\mathbb{C}} d^2 y \ |x-y|^{-4j} \Phi_{1-j} (y;z)
\end{eqnarray}
where the reflection amplitude is:
\begin{eqnarray}
\mathcal{R}(j) &=& (1-2j)
\frac{\Gamma \left( 1-\frac{1-2j}{k} \right)}{\Gamma \left(
    1+\frac{1-2j}{k} \right)} \left( \frac{\Gamma \left( 1+\frac{1}{k}
    \right)}{\Gamma(1-\frac{1}{k})} \right)^{1-2j}
\end{eqnarray}
The operator product expansions between primary fields and currents are of the form:
\begin{eqnarray}
\label{opeJPhi}
J^a (z) \Phi_j (x;w) &=& - \frac{D^a \Phi_j (x;w)}{z-w}  \\
D^3 = x \partial_x + j \ , \hspace{.8cm} D^+ &=& x^2 \partial_x + 2 j x \ ,
\hspace{.8cm} D^- = \partial_x \nonumber
\end{eqnarray}
The two-point function of primary fields is:
\begin{eqnarray}
\langle \Phi_{j_1} (x_1;z_1) \Phi_{j_2} (x_2;z_2) \rangle =
  \frac{A(j_1)}{|z_{12}|^{4h_{j_1}}} \left( \delta^2(x_{12})
  \delta(1-j_1-j_2) + \frac{\mathcal{R}(j_1)}{\pi}
  \frac{\delta(j_1-j_2)}{|x_{12}|^{4j_1}} \right)
\end{eqnarray}
with $A(j) = -\frac{\pi^2}{(2j-1)^2}$.  The expression for the
three-point function is
\cite{Teschner:1997ft,Ishibashi:2000fn,Giribet:2001ft}:
\begin{eqnarray}
\langle \Phi_{j_1} (x_1;z_1) \Phi_{j_2} (x_2;z_2) \Phi_{j_3}
  (x_3;z_3) \rangle = \prod_{1 \leq k<l \leq 3} \frac{1}{|z_{kl}|^{2h_{kl}}}
  \frac{1}{|x_{kl}|^{2j_{kl}}} D(j_1,j_2,j_3)
\end{eqnarray}
where $h_{kl} = h_k+h_l-h_m$, with $m \in \{1,2,3\}$ and $m \neq k,l$
(same for $j_{kl}$), and:
\begin{eqnarray}
D(j_1,j_2,j_3) = \frac{\pi}{2k} \left( k^{1/k}
  \frac{\Gamma \left(1+\frac{1}{k} \right)}{\Gamma \left(1-\frac{1}{k}
  \right)} \right)^{1-j_1-j_2-j_3}
\frac{\Upsilon (b) \Upsilon (2bj_1) \Upsilon (2bj_2)
  \Upsilon (2bj_3)}{\Upsilon (b(j_1+j_2+j_3-1)) \Upsilon (bj_{12})
  \Upsilon (bj_{13}) \Upsilon (bj_{23})} \nonumber
\end{eqnarray}
where the function $\Upsilon$ is defined in Appendix \ref{useful}
equation (\ref{defups}).

Note that the coefficient $D$ satisfies the following relation,
imposed by the reflection property of primary fields:
\begin{eqnarray}
\frac{D(j_1,j_2,j_3)}{D(j_1,j_2,1-j_3)} = \mathcal{R}(j_3)
\gamma(1-2j_3) \gamma (j_{13}) \gamma (j_{23})
\end{eqnarray}
The operator product expansion between two primary fields can be
deduced from the two- and three-point functions. For $z_1 \rightarrow
z_2$, one has:
\begin{eqnarray}
\label{opej}
\Phi_{j_1} (x_1;z_1) \Phi_{j_2} (x_2;z_2) \sim \int_0^{\infty}
d\lambda_3 \ \frac{1}{|z_{12}|^{2h_{12}}} \int_{\mathbb{C}} d^2 x_3
\prod_{1 \leq k<l \leq 3} \frac{1}{|x_{kl}|^{2j_{kl}}}
\frac{D(j_1,j_2,j_3)}{A(j_3)} \Phi_{1-j_3} (x_3;z_3) \nonumber \\
\end{eqnarray}
The fusion coefficient ${\mathcal{N}_{j_1, j_2}}^{j_3}$ is defined to
be one if $\Phi_{1-j_3}$ appears with a non-zero factor in the
operator product expansion of $\Phi_{j_1}$ and $\Phi_{j_2}$, and zero
otherwise.

\begin{figure}
\centerline{\psfig{figure=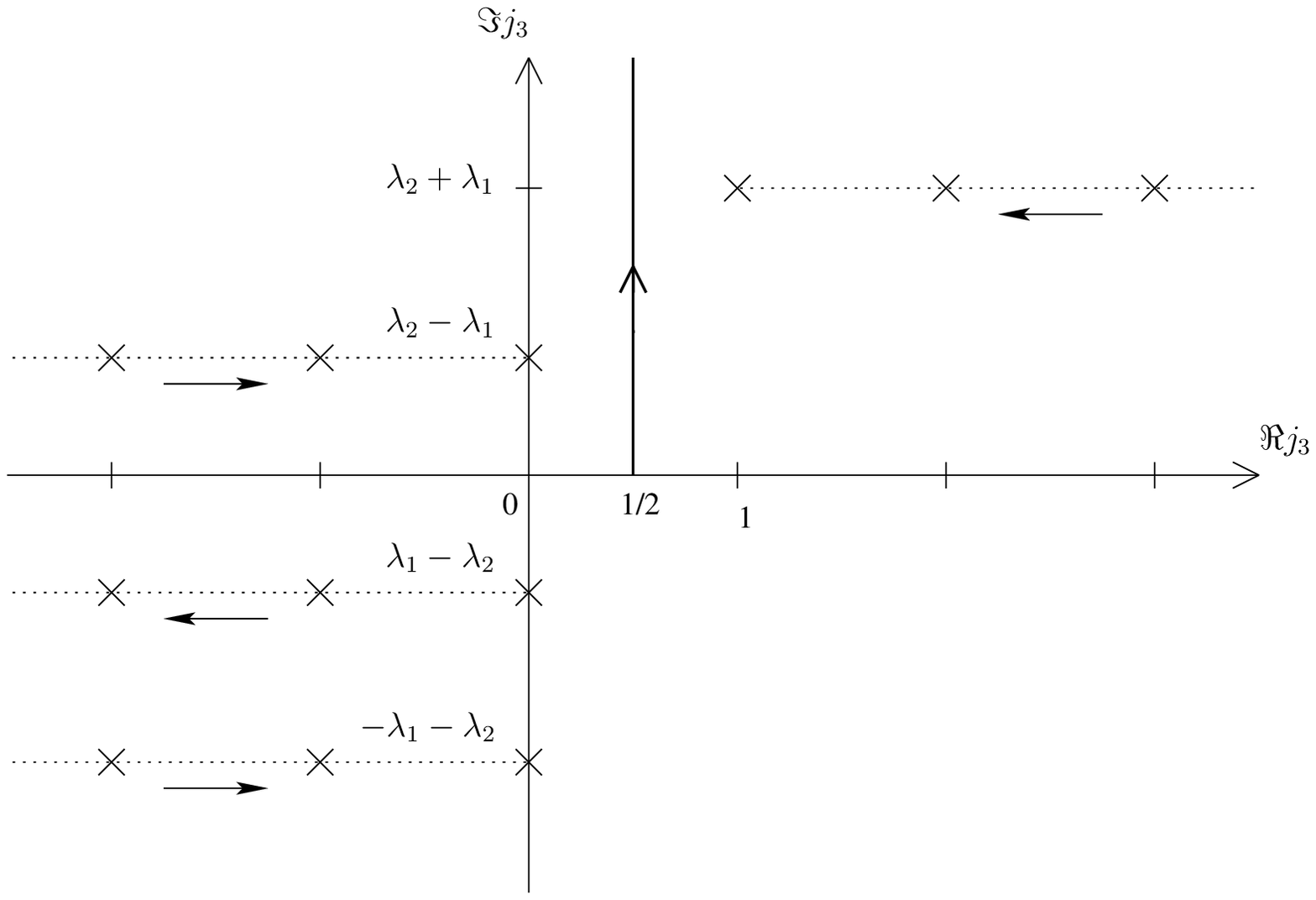,height=6cm}}
\caption{Integration contour and poles for the operator product
  expansion of a degenerate field and a non-degenerate field.}
\label{fusion1h3}
\end{figure}

\begin{figure}
\centerline{\psfig{figure=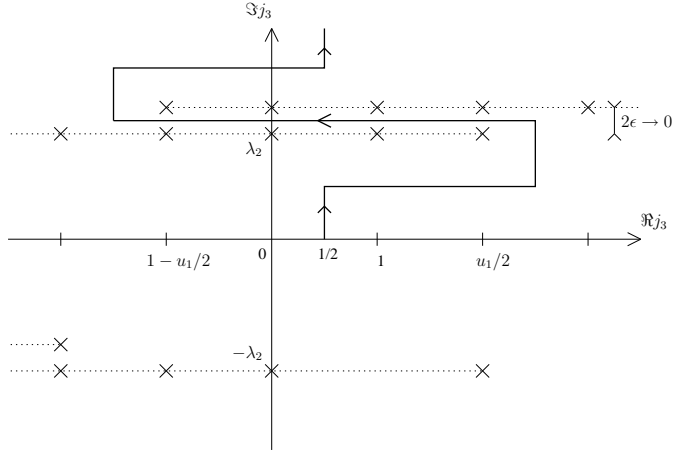,height=6cm}}
\caption{Integration contour and poles for the operator product
  expansion of a degenerate field and a non-degenerate field.}
\label{fusion2h3}
\end{figure}

The above two- and three-point functions and the operator product
expansion were given for continuous representations, for which all
factors are well-defined. Considering degenerate representations
requires a little bit more work, as we will see in the following. In
order to find the fusion  for a  degenerate and  a non-degenerate
representation, one deforms the contour of integration of the operator
product expansion \cite{Satoh:2001bi,Teschner:1999ug}, from the
initial situation shown in figure~\ref{fusion1h3} for which
$j_1=\frac{1}{2}+\imath \lambda_1$, $j_2=\frac{1}{2}+\imath
\lambda_2$, $\lambda_1, \lambda_2 \in \mathbb{R}_+$, to the case
$\lambda_1 = \imath \frac{u_1}{2} + \epsilon$, with $u_1 \in
\mathbb{N}^*$ and $\epsilon$ a positive infinitesimal number, shown in
figure~\ref{fusion2h3}. The figures show the poles (pictured  as
crosses) of $D(j_1,j_2,j_3)$ in  the $j_3$ complex plane. One  should
also  take into account  zeros in the numerator of $D(j_1,j_2,j_3)$
that  appear in  the limit $\epsilon \rightarrow 0$. Arrows in
figure~\ref{fusion1h3} indicate in which direction poles move when
$\Im \lambda_1$ increases from $0$ up to $\frac{u_1}{2}$.  For
$\epsilon \rightarrow 0$, the contour of integration  is pinched
between some poles (note that when two of these poles merge, there is
an extra zero factor coming from $\Upsilon (2 j_1  b)$ which make the
total residue non-zero). Then, we must pull the integration contour
over the poles, and the integral is transformed  into  a sum over all
non-zero residues (note that there is no other contribution to the
operator product expansion, because in the limit $\epsilon \rightarrow
0$, $D(j_1,j_2,j_3)=0$ except at its poles). The final result is
consistent with the expectation for the fusion of degenerate and
non-degenerate representations in $H_3^+$ and it is given in the bulk
of the paper in equation $(\ref{fusionnondegh3})$ (where the notation
is $u = 2J+1$).


\begin{figure}
\centerline{\psfig{figure=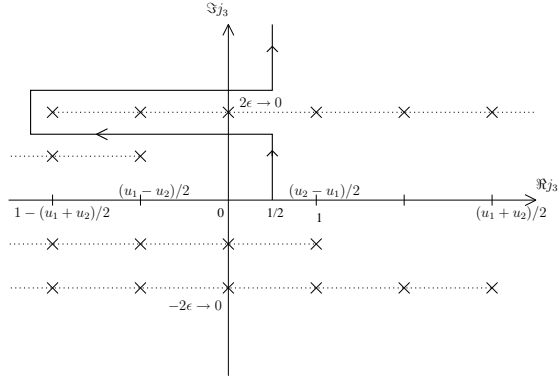,height=5cm}}
\caption{Integration contour and poles for the operator product
  expansion of two degenerate fields.}
\label{fusion3h3}
\end{figure}

For the fusion of two degenerate representations, one starts again
from figure~\ref{fusion1h3}, but then deforms the integration contour
to $\lambda_1 \rightarrow \imath \frac{u_1}{2} + \frac{\epsilon}{2}$
and $\lambda_2 \rightarrow \imath \frac{u_2}{2} + \frac{3
  \epsilon}{2}$, as shown in figure~\ref{fusion3h3} in the case $u_2
\geq u_1$. Just as before, the contour of integration is pinched
between some poles, which are going to contribute to the fusion i.e.
to a sum over residues. The result is given in the bulk of the paper
in equation $(\ref{fusdegh3})$.


\section{Fusion in bosonic Liouville theory}
In this appendix we consider the fusion of degenerate (and
non-degenerate) representations in bosonic Liouville theory, 
as obtained from analytically continuing three-point functions for 
non-degenerate unitary representations
\cite{Dorn:1992xw,Dorn:1992at,Zamolodchikov:1996xm} (see 
also  \cite{Teschner:1999ug}).
Our analysis is more complete than those in the literature, and still
it contains some puzzling features.

\label{appliouville}

Liouville theory is a field theory classically described by the
lagrangian:
\begin{eqnarray}
\mathcal{L} = \frac{1}{4 \pi} (\partial \Phi)^2 + \mu e^{2 b \Phi}
\end{eqnarray}
where $\Phi$ is the Liouville field, $b$ is the dimensionless coupling
constant which is strictly positive and such that $b^2$ is non-rational, and
$\mu$ is a scale parameter called the cosmological constant.

The theory is conformally invariant with central charge:
\begin{eqnarray}
c = 1+ 6 Q^2 \ , \hspace{1cm} Q = b + b^{-1}
\end{eqnarray}
Primary fields are of the form $V_{\alpha} = e^{2 \alpha \Phi}$, and
have conformal weight:
\begin{eqnarray}
h_{\alpha} = \alpha (Q-\alpha)
\end{eqnarray}
Note that primaries $V_{\alpha}$ and $V_{Q-\alpha}$ have the same
conformal weight and are closely related. More precisely, the
Liouville reflection amplitude reads \cite{Zamolodchikov:1996xm}:
\begin{eqnarray}
\mathcal{R}_L (\alpha) = - \left( \pi \mu \gamma (b^2)
\right)^{(Q-2\alpha)/b} \frac{\Gamma(1-(Q-2\alpha) b)
  \Gamma \left(1-\left(\frac{Q-2\alpha}{b}\right) \right)}{\Gamma(1+(Q-2\alpha) b)
  \Gamma \left(1+\left(\frac{Q-2\alpha}{b}\right) \right)}
\end{eqnarray}
and allows us to write $V_{\alpha} = \mathcal{R}_L (\alpha)
V_{Q-\alpha}$, a relation which holds in any correlation function.

Physical (unitary) representations are obtained for $2 \alpha = Q +
\imath s$ with $s \in \mathbb{R}$, which can be restricted to $s \in
\mathbb{R}_+$ because of the reflection. They are
non-degenerate. There exist degenerate representations as well,
characterized by $2 \alpha = \frac{1-m}{b} + (1-n) b$ with $m, n \in
\mathbb{N}^*$ (see \cite{Zamolodchikov:2001ah}), but they do not
belong to the unitary conformal field theory spectrum.

The two- and three-point function are given by
\cite{Dorn:1992xw,Dorn:1992at,Zamolodchikov:1996xm}:
\begin{eqnarray}
\langle \Phi_{\alpha_1} (z_1) \Phi_{\alpha_2}
  (z_2) \rangle = \frac{2 \pi}{|z_{12}|^{4h_1}} (
  \delta(Q-\alpha_1-\alpha_2) + \mathcal{R}_L (\alpha_1)
  \delta(\alpha_1-\alpha_2) )
\end{eqnarray}
\begin{eqnarray}
\langle \Phi_{\alpha_1} (z_1) \Phi_{\alpha_2} (z_2) \Phi_{\alpha_3}
  (z_3) \rangle =  \prod_{1 \leq k<l \leq 3} \frac{1}{|z_{kl}|^{2h_{kl}}}
  \ C(\alpha_1,\alpha_2,\alpha_3)
\end{eqnarray}
where $h_k = h_{\alpha_k}$, $h_{kl} = h_k+h_l-h_m$ with $m \in
\{1,2,3\}$ and $m \neq k,l$, and:
\begin{eqnarray}
\label{c123}
C(\alpha_1,\alpha_2,\alpha_3) =
  \beta^{(Q-\alpha_1-\alpha_2-\alpha_3)/b} \frac{\Upsilon(b)
  \Upsilon(2 \alpha_1) \Upsilon(2 \alpha_2) \Upsilon(2
  \alpha_3)}{\Upsilon(\alpha_1+\alpha_2+\alpha_3-Q)
  \Upsilon(\alpha_{12}) \Upsilon(\alpha_{13}) \Upsilon(\alpha_{23})}
\end{eqnarray}
where $\beta = \pi \mu \gamma (b^2) b^{2-2b^2}$, $\alpha_{kl} =
\alpha_k+\alpha_l-\alpha_m$ with $m \in \{1,2,3\}$ and $m \neq k,l$,
and $\Upsilon$ is a function defined in Appendix \ref{useful}.

The operator product expansion of non-degenerate fields is given by
($z_1 \rightarrow z_2$):
\begin{eqnarray}
\Phi_{\alpha_1} (z_1) \Phi_{\alpha_2} (z_2) \sim
\int d\alpha_3 \frac{1}{|z_{12}|^{2h_{12}}}
\frac{C(\alpha_1,\alpha_2,\alpha_3)}{4 \pi} \Phi_{Q-\alpha_3} (z_2)
\end{eqnarray}
where $\int d\alpha_3 = \frac{1}{2} \int_{\mathbb{R}} d s_3 =
\int_{\mathbb{R}_+} d s_3$. The operator product expansion is
consistent with the two- and three-point function.

The fusion coefficients ${\mathcal{N}_{\alpha_1,\alpha_2}}^{\alpha_3}$
is defined to be one if $\Phi_{Q-\alpha_3}$ appears with a non-zero
factor in the operator product expansion of $\Phi_{\alpha_1}$ and
$\Phi_{\alpha_2}$, and zero otherwise.

\begin{figure}
\centerline{\psfig{figure=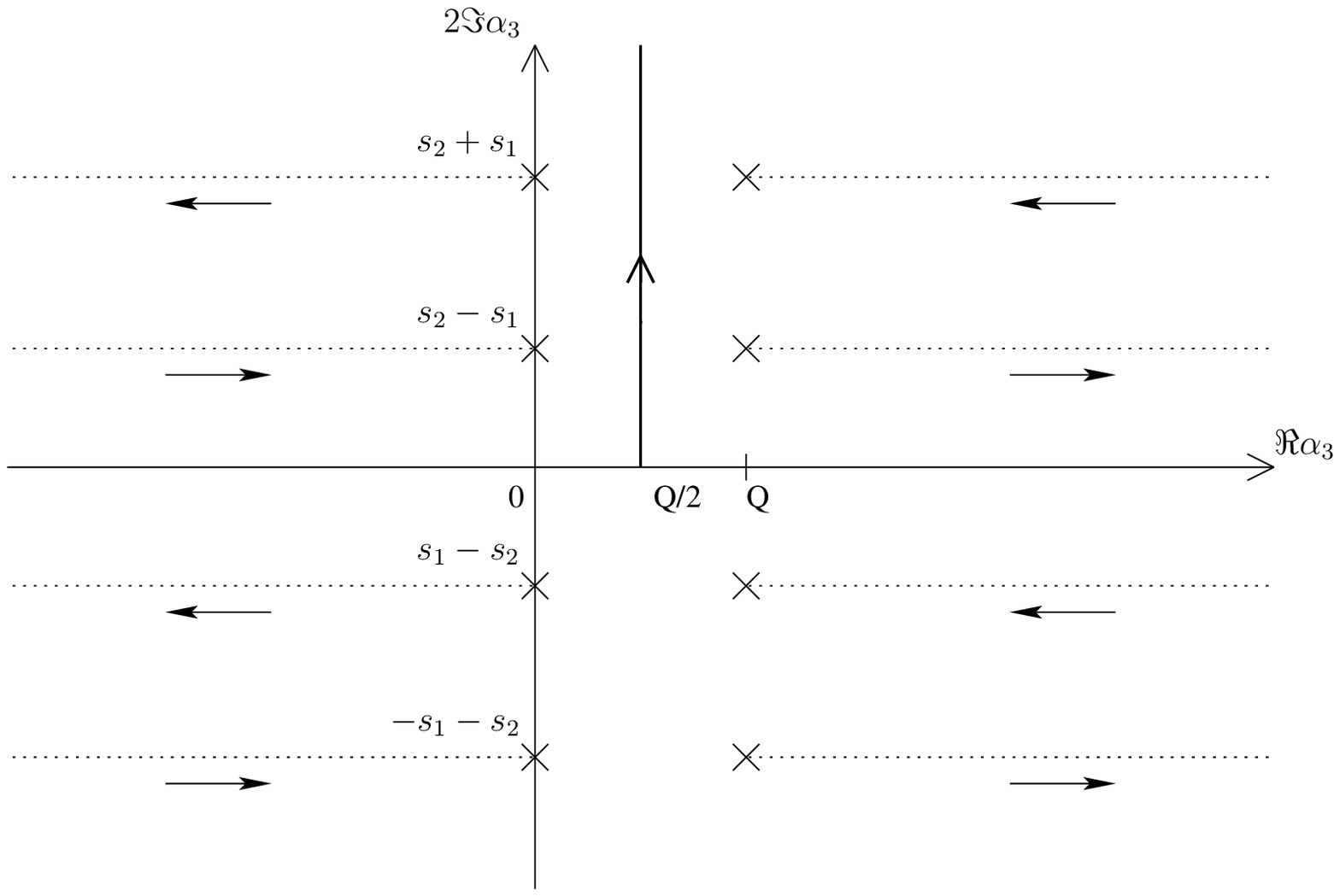,height=6cm}}
\caption{Integration contour and poles for the operator product
  expansion of two non-degenerate fields.}
\label{fusion1liou}
\end{figure}

\begin{figure}
\centerline{\psfig{figure=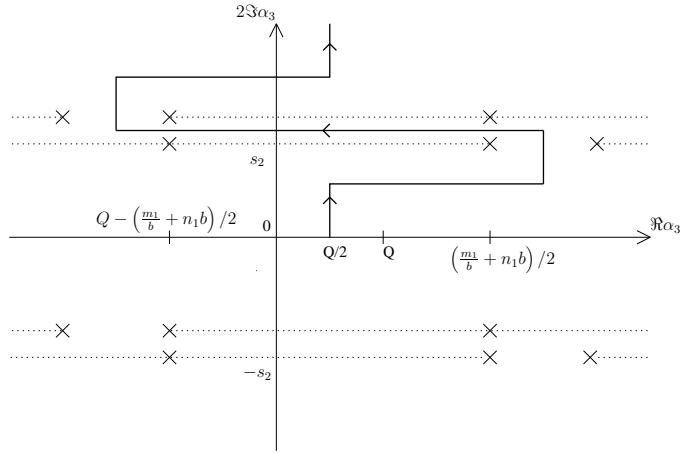,height=6cm}}
\caption{Integration contour and poles for the operator product
  expansion of a degenerate field and a non-degenerate field.}
\label{fusion2liou}
\end{figure}

The procedure to follow in order to find the fusion for degenerate
fields is the same as for $H_3^+$. Figure~\ref{fusion1liou} shows
poles of $C(\alpha_1,\alpha_2,\alpha_3)$ in the $\alpha_3$ complex
plane (as opposed to the $H_3^+$ case where there are less poles, only
some poles are pictured here as crosses, while other poles are on the
dotted half-lines), for $2 \alpha_1 = Q + \imath s_1$ and $2 \alpha_2
= Q + \imath s_2$, and the direction in which these poles move when
$\Im s_1$ increases from $0$ up to $\frac{m_1}{b} + n_1 b$, with $m_1,
n_1 \in \mathbb{N}^*$. Figure~\ref{fusion2liou} shows poles for $s_1 =
\imath \left( \frac{m_1}{b} + n_1 b \right) + \epsilon$, with
$\epsilon$ a positive infinitesimal number. The poles which pinch the
operator product expansion contour integral give us the fusion
coefficients of a degenerate representation with a non-degenerate
one. The result was given in the bulk of the paper in equation
(\ref{fuslioundeg}).

We further remark that the fusion of a degenerate with any non-degenerate
field with an imaginary part to its Liouville momentum yields the same fusion
relations, quoted in the bulk of the paper. Again, this can be demonstrated 
on the basis of analytically continuing the operator product expansion in
Liouville theory.
We conclude that for any continuous limit of fusion in
Liouville theory, the fusion of
degenerate fields too, will satisfy the expected fusion relations.
\footnote{We would like to thank an
  anonymous referee for a useful comment on this point.}
This then gives the canonical fusion rules for degenerate fields, after using
the standard symmetry argument (see e.g. \cite{DiFrancesco:1997nk}). 

\begin{figure}
\centerline{\psfig{figure=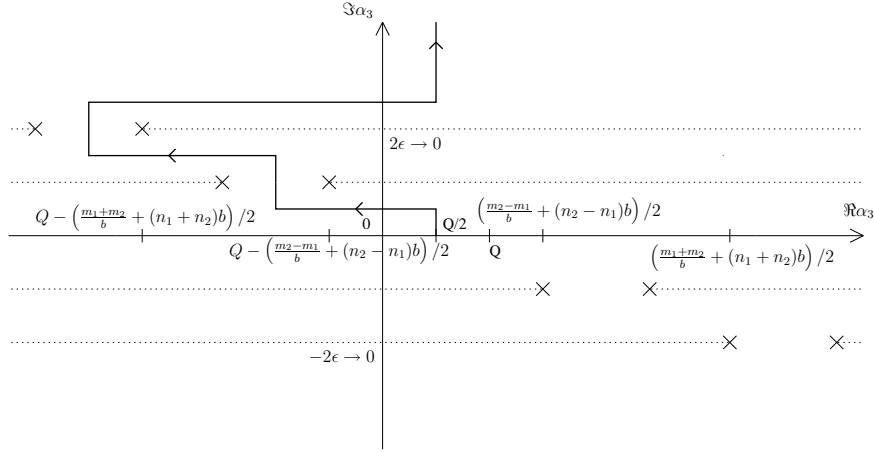,height=6cm}}
\caption{Integration contour and poles for the operator product
  expansion of a degenerate field and another degenerate field.}
\label{fusion3liou}
\end{figure}

We note that there is an interesting side-remark to be made here. If we consider
recuperating degenerate fusion by analytic continuation in the three-point
functions (instead of analytic continuation in the fusion coefficients), we
find an interesting subtlety.
Figure~\ref{fusion3liou} shows poles for the case of two degenerate
representations, i.e when one has $s_1 = \imath \left( \frac{m_1}{b} +
  n_1 b \right) + \epsilon$ and $s_2 = \imath \left( \frac{m_2}{b} +
  n_2 b \right) + 3 \epsilon$. The fusion coefficients given by this
analysis were given in the bulk of the paper in equation
$(\ref{fusionliou2deg})$ for the case $m_2 \geq m_1+1$ and $n_2 \leq
n_1$. The result would be the same if we had $m_1 \geq m_2+1$ and $n_1
\leq n_2$. For the case $m_2 \geq m_1 +1$ and $n_2 \geq n_1+1$, or
equivalently $m_1 \geq m_2 +1$ and $n_1 \geq n_2+1$, the subtlety is
that poles (or rather triple zeroes compensated by four poles)
 appear in the segment $Q-\left( (m_2-m_1)/b + (n_2-n_1) b
\right)/2 \leq \alpha_3 \leq Q/2$. 

Thus, what we find is that, due to the fact that the three-point function is
not analytic in the momenta, and can blow up at certain points, the decoupling
of degenerate representations is not realized at particular analytically
continued momenta. This is not in contradiction with the decoupling of
degenerate representations from the unitary spectrum, nor is it in
contradiction with the standard argument for decoupling (see
e.g. \cite{DiFrancesco:1997nk}), 
which assumes a finite three-point function. In fact, it is not surprising
that the formal limit of Liouville theory does not give rise to standard
decoupling -- it would otherwise provide a rational conformal field theory
of degenerate fields at any value of the central charge. 

In conclusion, we recuperate the standard
fusion relations for degenerate representations (which follow from
decoupling), by
 considering the
analytic continuation in the fusion coefficients for degenerate/non-degenerate
fusion.


\section{Useful formulas}
\label{useful}
It is rather standard to define the $\gamma$ function as:
\begin{eqnarray}
\gamma (z) = \frac{\Gamma(z)}{\Gamma(1-z)}
\end{eqnarray}
The $\Upsilon$ function is defined on the strip $0 < \Re (x) < Q$ by the
following integral representation:
\begin{eqnarray}
\ln{\left( \Upsilon (x) \right)} = \int_0^{\infty} \frac{dt}{t} \left[
  \left( \frac{Q}{2} - x \right)^2 e^{-t} - \frac{  sh^2\left( \left(
  \frac{Q}{2} - x \right) \frac{t}{2} \right)  }{  sh \left( \frac{b
  t}{2} \right) sh \left( \frac{t}{2 b} \right)  } \right]
\label{defups}
\end{eqnarray}
where $Q = b + 1/b$ and $b \in \mathbb{R}^*$. The function $\Upsilon$
can be extended to the whole complex plane, thanks to the relations:
\begin{eqnarray}
\Upsilon (x+b) &=& \gamma (bx) b^{1-2bx} \Upsilon (x) \nonumber \\
\Upsilon (x+1/b) &=& \gamma (x/b) b^{2x/b-1} \Upsilon (x)
\end{eqnarray}
The function $\Upsilon$ is entire in the variable $x$ with zeroes at $x
= - x_{m,n}$ and at $x = Q + x_{m,n}$, with $x_{m,n} = m/b+nb$ and $m,n \in
\mathbb{N}$. Other relations satisfied by the function $\Upsilon$ are:
\begin{eqnarray}
\Upsilon (Q-x) = \Upsilon (x) \ , \hspace{1cm} \Upsilon (Q/2) = 1 \ ,
\hspace{1cm} \Upsilon'(0) &=& \Upsilon (b)
\end{eqnarray}
We also made use of the Dotsenko-Fateev integral \cite{Dotsenko:1984nm} given by:
\begin{eqnarray}
I_n (\alpha, \beta, \rho) &=& \int \prod_{i=1}^n d^2 y_i |y_i|^{2
  \alpha} |1-y_i|^{2 \beta} \prod_{i<j} |y_i - y_j|^{4 \rho} \\
&=& \pi^n n! \prod_{l=0}^{n-1} \frac{ \gamma((l+1) \rho)}{\gamma (\rho)}
  \frac{\gamma (1 + \alpha + l \rho) \gamma (1 + \beta + l
  \rho)}{\gamma (2 + \alpha + \beta + (n-1+l) \rho)} \nonumber
\end{eqnarray}
And the following integrals were equally useful:
\begin{eqnarray}
\int_{\mathbb{C}} d^2 y |x-y|^{-4j-4} y^{j-m} \bar{y}^{j-\bar{m}}
&=& \pi \frac{\Gamma (1+j-m) \Gamma (1+j+\bar{m})}{\Gamma (-j-m) \Gamma
  (-j+\bar{m})} \\
& & \times \frac{\Gamma (-2j-1)}{\Gamma (2j+2)} x^{-j-1-m}
\bar{x}^{-j-1-\bar{m}} \nonumber
\end{eqnarray}
with $m-\bar{m} \in \mathbb{Z}$ and $\Re j > -1$, and, for $n, m \in
\mathbb{Z}$ (see \cite{Giveon:2001up}):
\begin{eqnarray}
\int_{\mathbb{C}} d^2x |x|^{2a} x^n |1-x|^{2b} (1-x)^m = \pi \frac{\Gamma(a+n+1)
  \Gamma(b+m+1) \Gamma(-a-b-1)}{\Gamma(-a) \Gamma(-b)
  \Gamma(a+b+n+m+2)}
\end{eqnarray}


\newpage

\begin{thebibliography}{99}

\bibitem{DiFrancesco:1997nk}
P.~Di Francesco, P.~Mathieu and D.~Senechal, Conformal field theory,
Springer New-York 1997.

\bibitem{Cappelli:1987xt}
A.~Cappelli, C.~Itzykson and J.~B.~Zuber,
Commun.\ Math.\ Phys.\  {\bf 113}, 1 (1987).

\bibitem{Gannon:2001uy}
  T.~Gannon,
  arXiv:math.qa/0103044.

\bibitem{Verlinde:1988sn}
  E.~Verlinde,
  Nucl.\ Phys.\ B {\bf 300} (1988) 360.

\bibitem{Zamolodchikov:2001ah}
  A.~B.~Zamolodchikov and A.~B.~Zamolodchikov,
  arXiv:hep-th/0101152.

\bibitem{Fateev:2000ik}
  V.~Fateev, A.~B.~Zamolodchikov and A.~B.~Zamolodchikov,
  arXiv:hep-th/0001012.

\bibitem{Teschner:2000md}
  J.~Teschner,
  arXiv:hep-th/0009138.

\bibitem{Eguchi:2003ik}
  T.~Eguchi and Y.~Sugawara,
  JHEP {\bf 0401}, 025 (2004)
  [arXiv:hep-th/0311141].

\bibitem{Israel:2004jt}
  D.~Israel, A.~Pakman and J.~Troost,
  Nucl.\ Phys.\ B {\bf 710}, 529 (2005)
  [arXiv:hep-th/0405259].

\bibitem{Fotopoulos:2004ut}
  A.~Fotopoulos, V.~Niarchos and N.~Prezas,
  Nucl.\ Phys.\ B {\bf 710}, 309 (2005)
  [arXiv:hep-th/0406017].

\bibitem{Ahn:2004qb}
C.~Ahn, M.~Stanishkov and M.~Yamamoto,
JHEP {\bf 0407} (2004) 057
[arXiv:hep-th/0405274].

\bibitem{Cardy:1989ir}
J.~L.~Cardy,
Nucl.\ Phys.\ B {\bf 324} (1989) 581.

\bibitem{Fuchs:1993et}
  J.~Fuchs,
  Fortsch.\ Phys.\  {\bf 42} (1994) 1
  [arXiv:hep-th/9306162].

\bibitem{Moore:1988qv}
  G.~W.~Moore and N.~Seiberg,
  Commun.\ Math.\ Phys.\  {\bf 123} (1989) 177.

\bibitem{Huang:2004dg}
Y.~Z.~Huang,
Proc.\ Nat.\ Acad.\ Sci.\  {\bf 102}, 5352 (2005)
[arXiv:math.qa/0412261].

\bibitem{Gannon:2003de}
  T.~Gannon,
  Nucl.\ Phys.\ B {\bf 670} (2003) 335
  [arXiv:hep-th/0305070].

\bibitem{Awata:1992sm}
  H.~Awata and Y.~Yamada,
  Mod.\ Phys.\ Lett.\ A {\bf 7}, 1185 (1992).

\bibitem{Ramgoolam:1993gt}
  S.~Ramgoolam,
  arXiv:hep-th/9301121.

\bibitem{Feigin:1993dt}
  B.~Feigin and F.~Malikov,
  Lett.\ Math.\ Phys.\  {\bf 31}, 315 (1994)
  [arXiv:hep-th/9310004].

\bibitem{Gaberdiel:2001ny}
  M.~R.~Gaberdiel,
  Nucl.\ Phys.\ B {\bf 618} (2001) 407
  [arXiv:hep-th/0105046].
 
\bibitem{Lesage:2002ch}
  F.~Lesage, P.~Mathieu, J.~Rasmussen and H.~Saleur,
  Nucl.\ Phys.\ B {\bf 647}, 363 (2002)
  [arXiv:hep-th/0207201].

\bibitem{Zamolodchikov:1996xm}
  A.~B.~Zamolodchikov and A.~B.~Zamolodchikov,
  {\it Prepared for 2nd International Sakharov Conference on Physics,
    Moscow, Russia, 20-23 May 1996}

\bibitem{Ponsot:2001gt}
  B.~Ponsot, V.~Schomerus and J.~Teschner,
  JHEP {\bf 0202} (2002) 016
  [arXiv:hep-th/0112198].

\bibitem{Giveon:2001uq}
A.~Giveon, D.~Kutasov and A.~Schwimmer,
Nucl.\ Phys.\ B {\bf 615} (2001) 133
[arXiv:hep-th/0106005].

\bibitem{Kato:1991zz}
  M.~Kato and Y.~Yamada,
  Prog.\ Theor.\ Phys.\ Suppl.\  {\bf 110} (1992) 291.

\bibitem{Miki:1989ri}
  K.~Miki,
  Int.\ J.\ Mod.\ Phys.\ A {\bf 5} (1990) 1293.

\bibitem{Ribault:2003ss}
S.~Ribault and V.~Schomerus,
JHEP {\bf 0402} (2004) 019
[arXiv:hep-th/0310024].

\bibitem{Israel:2004xj}
  D.~Israel, A.~Pakman and J.~Troost,
  JHEP {\bf 0404}, 045 (2004)
  [arXiv:hep-th/0402085].


\bibitem{Hosomichi:2004ph}
  K.~Hosomichi,
  arXiv:hep-th/0408172.


\bibitem{Maldacena:2000kv}
  J.~M.~Maldacena, H.~Ooguri and J.~Son,
  J.\ Math.\ Phys.\  {\bf 42}, 2961 (2001)
  [arXiv:hep-th/0005183].

\bibitem{Hanany:2002ev}
  A.~Hanany, N.~Prezas and J.~Troost,
  JHEP {\bf 0204}, 014 (2002)
  [arXiv:hep-th/0202129].

\bibitem{Seiberg:2003nm}
  N.~Seiberg and D.~Shih,
  JHEP {\bf 0402} (2004) 021
  [arXiv:hep-th/0312170].

\bibitem{D'Appollonio:2004pm}
  G.~D'Appollonio and E.~Kiritsis,
  Nucl.\ Phys.\ B {\bf 712}, 433 (2005)
  [arXiv:hep-th/0410269].

\bibitem{Belavin:1984vu}
  A.~A.~Belavin, A.~M.~Polyakov and A.~B.~Zamolodchikov,
  Nucl.\ Phys.\ B {\bf 241} (1984) 333.

\bibitem{Gepner:1986wi}
  D.~Gepner and E.~Witten,
  Nucl.\ Phys.\ B {\bf 278}, 493 (1986).

\bibitem{Gaberdiel:1993td}
  M.~Gaberdiel,
  Int.\ J.\ Mod.\ Phys.\ A {\bf 9} (1994) 4619
  [arXiv:hep-th/9307183].

\bibitem{Fuchs:1997af}
  J.~Fuchs,
  arXiv:hep-th/9702194.

\bibitem{Teschner:1997ft}
  J.~Teschner,
  Nucl.\ Phys.\ B {\bf 546} (1999) 390
  [arXiv:hep-th/9712256].

\bibitem{Hosomichi:2001fm}
K.~Hosomichi and Y.~Satoh,
Mod.\ Phys.\ Lett.\ A {\bf 17} (2002) 683
[arXiv:hep-th/0105283].

\bibitem{Satoh:2001bi}
Y.~Satoh,
Nucl.\ Phys.\ B {\bf 629} (2002) 188
[arXiv:hep-th/0109059].

\bibitem{Maldacena:2001km}
J.~M.~Maldacena and H.~Ooguri,
Phys.\ Rev.\ D {\bf 65} (2002) 106006
[arXiv:hep-th/0111180].

\bibitem{Ishibashi:2000fn}
N.~Ishibashi, K.~Okuyama and Y.~Satoh,
Nucl.\ Phys.\ B {\bf 588} (2000) 149
[arXiv:hep-th/0005152].

\bibitem{Giribet:2001ft}
G.~Giribet and C.~Nunez,
JHEP {\bf 0106} (2001) 010
[arXiv:hep-th/0105200].

\bibitem{Teschner:1999ug}
  J.~Teschner,
  Nucl.\ Phys.\ B {\bf 571} (2000) 555
  [arXiv:hep-th/9906215].

\bibitem{Dorn:1992xw}
  H.~Dorn and H.~J.~Otto,
  arXiv:hep-th/9212004.

\bibitem{Dorn:1992at}
  H.~Dorn and H.~J.~Otto,
  Phys.\ Lett.\ B {\bf 291}, 39 (1992)
  [arXiv:hep-th/9206053].

\bibitem{Dotsenko:1984nm}
  V.~S.~Dotsenko and V.~A.~Fateev,
  Nucl.\ Phys.\ B {\bf 240} (1984) 312.
  Nucl.\ Phys.\ B {\bf 251} (1985) 691.

\bibitem{Giveon:2001up}
  A.~Giveon and D.~Kutasov,
  Nucl.\ Phys.\ B {\bf 621} (2002) 303
  [arXiv:hep-th/0106004].


\end{thebibliography}
\end{document}